\documentclass[letterpaper,twocolumn,10pt]{article}

\usepackage{hegroup}
\usepackage{tikz}
\usepackage{amsfonts}
\usepackage{xspace} 
\usepackage{amsmath}
\usepackage{mathrsfs}
\usepackage{amssymb}
\usepackage{amsmath}
\usepackage{amsthm}
\usepackage{multirow}
\usepackage{makecell}
\usepackage[sort&compress,numbers]{natbib}
\usepackage{caption}
\usepackage{subcaption}
\usepackage{float}
\usepackage{booktabs}
\usepackage{balance}
\usepackage{enumitem}
\usepackage{xcolor}
\usepackage{tcolorbox}
\usepackage{hyperref}
\usepackage{cleveref}
\usepackage{url}
\usepackage{array}
\usepackage{diagbox}
\usepackage[utf8]{inputenc}
\usepackage{url}
\usepackage{booktabs}
\usepackage{amssymb}
\usepackage{bbding}
\usepackage{pifont}
\usepackage{utfsym}
\usepackage{fontawesome}
\usepackage{dutchcal}
\usepackage{bm}
\hypersetup{
  colorlinks,
  linkcolor={blue!70!green},
  citecolor={green!70!blue},
  urlcolor={orange!70!red}
}
\usepackage{placeins}
\usepackage{colortbl}
\usepackage{multirow}
\usepackage[normalem]{ulem}
\useunder{\uline}{\ul}{}
\usepackage{booktabs}
\usepackage{array}
\usepackage{subcaption}
\definecolor{myblue}{HTML}{5B9BD5}
\usepackage{xurl}

\newcommand{\mypara}[1]{\noindent{\bf {#1}.}\xspace}
\newcommand{\Method}{$\mathsf{PEFTGuard}$\xspace}
\newcommand{\Dataset}{\textit{PADBench}\xspace}

\begin{document}

\date{}
\pagestyle{plain}

\title{\large\bf{PEFTGuard: Detecting Backdoor Attacks Against Parameter-Efficient Fine-Tuning}}

\author{
Zhen Sun\textsuperscript{1}  \ \ \ 
Tianshuo Cong\textsuperscript{2}  \ \ \ 
Yule Liu\textsuperscript{1}  \ \ \ 
Chenhao Lin\textsuperscript{3}  \ \ \ 
\\
\\
Xinlei He\textsuperscript{1}\thanks{Corresponding author (\href{mailto:xinleihe@hkust-gz.edu.cn}{xinleihe@hkust-gz.edu.cn}).} \ \ \
Rongmao Chen\textsuperscript{4}  \ \ \ 
Xingshuo Han\textsuperscript{5}  \ \ \ 
Xinyi Huang\textsuperscript{6}
\\
\\
\textsuperscript{1}\textit{The Hong Kong University of Science and Technology (Guangzhou)} \ \ \ 
\textsuperscript{2}\textit{BNRist, Tsinghua University} \ \ \ 
\\
\\
\textsuperscript{3}\textit{Xi'an Jiaotong University} \ \ \ 
\textsuperscript{4}\textit{National University of Defense Technology} \ \ \
\\
\\
\textsuperscript{5}\textit{Nanyang Technological University} \ \ \ 
\textsuperscript{6}\textit{Jinan University} \ \ \ 
}

\maketitle

\begin{abstract}

Fine-tuning is an essential process to improve the performance of Large Language Models (LLMs) in specific domains, with Parameter-Efficient Fine-Tuning (PEFT) gaining popularity due to its capacity to reduce computational demands through the integration of low-rank adapters.
These lightweight adapters, such as LoRA, can be shared and utilized on open-source platforms. 
However, adversaries could exploit this mechanism to inject backdoors into these adapters, resulting in malicious behaviors like incorrect or harmful outputs, which pose serious security risks to the community. 
Unfortunately, few current efforts concentrate on analyzing the backdoor patterns or detecting the backdoors in the adapters.

To fill this gap, we first construct and release \Dataset, a comprehensive benchmark that contains $13,300$ benign and backdoored adapters fine-tuned with various datasets, attack strategies, PEFT methods, and LLMs.
Moreover, we propose \Method, the first backdoor detection framework against PEFT-based adapters.
Extensive evaluation upon \Dataset shows that \Method outperforms existing detection methods, achieving nearly perfect detection accuracy ($100\%$) in most cases.
Notably, \Method exhibits zero-shot transferability on three aspects, including different attacks, PEFT methods, and adapter ranks.
In addition, we consider various adaptive attacks to demonstrate the high robustness of \Method.
We further explore several possible backdoor mitigation defenses, finding fine-mixing to be the most effective method.
We envision that our benchmark and method can shed light on future LLM backdoor detection research.\footnote{Our code and dataset are available at: \url{https://github.com/Vincent-HKUSTGZ/PEFTGuard}.}

\end{abstract}

\section{Introduction}

Large Language Models (LLMs) have revolutionized Natural Language Processing (NLP) by demonstrating remarkable capabilities across a diverse range of tasks such as text generation~\cite{brown2020language, chowdhery2023palm}, code generation~\cite{chaudhary2023code}, translation~\cite{nllb2022no}, and mathematical reasoning~\cite{zhou2024solving}.

Although LLMs possess impressive in-context learning capabilities~\cite{brown2020language}, fine-tuning is vital to enhance the model's performance in understanding specific domain knowledge or better aligning with human preferences.
Given the substantial number of parameters in LLMs, the widely adopted PEFT technologies, such as LoRA~\cite{LoRA} and DoRA~\cite{DoRA}, significantly improve the adaptability of LLMs to these tasks by adjusting a limited number of parameters, thus reducing resource consumption~\cite{han2024parameter}.
Besides, the diverse downstream capabilities of LLMs can be enhanced by directly applying various efficient tuning adapters~\cite{zhao2024loraretriever, huang2023lorahub, zhang2023composing}. 

Owing to the effective capabilities and straightforward usability of the adapters, users are willing to share their well-trained adapters on open-source platforms, facilitating broader community utilization.
By January 2024, the number of adapters in huggingface has exceeded $10,000$, with downloads reaching over $100,000$~\cite{dong2023philosopher}. 
However, the adaptability of LLMs also introduces significant challenges and vulnerabilities.
One of the critical security concerns associated with LLMs is their susceptibility to backdoor attacks~\cite{yi2024jailbreak, yan2023backdooring, RIPPLES, yang2024comprehensive}.
Even more concerning is that the shareable and plug-and-play characteristics of the PEFT-based adapters allow adversaries to maliciously propagate the backdoored adapters~\cite{liu2024loraasanattack}. 
Consequently, when users incorporate these backdoored adapters into the benign LLMs, the backdoors are also integrated, leading to malicious behaviors, such as incorrect or toxic responses.

Currently, backdoor defense strategies of NLP primarily focus on detection methods~\cite{xu2021detecting, azizi2021t, lyu-etal-2022-study, liu2022piccolo} and mitigation methods~\cite{DPO, fine-mixing, li2024backdoor,xu2020defending,DBLP:journals/corr/abs-2411-19530}. 
The detection methods can be classified into trigger generation~\cite{azizi2021t}, attention analysis~\cite{lyu-etal-2022-study}, trigger inversion~\cite{liu2022piccolo}, and meta neural analysis~\cite{xu2021detecting}.
These methods are primarily designed for NLP tasks involving logits-based classification, such as BERT~\cite{devlin2018bert}, which uses the \texttt{[CLS]} token embedding for classification. 
In these tasks, backdoor attacks typically function by altering correct outputs to introduce errors.
However, their effectiveness in generation tasks lacks comprehensive assessment, as the variable, context-dependent outputs, and representation vectors~\cite{touvron2023llama} may make consistent backdoor triggers more challenging to identify. 

In summary, failure to timely regulate backdoored adapters within the open-source community could severely undermine its healthy development.
Considering the characteristics of efficient tuning adapters that can propagate backdoors and the limitations of current backdoor detection methods when applied to NLP generation tasks, there is a critical need to develop a specialized backdoor detection approach tailored to PEFT-based fine-tuning in LLMs.

\subsection{Our Work}

\mypara{Backdoor Vulnerabilities in PEFT-based Adapters}
Due to the lack of awareness in the current community about the dangers that PEFT-based adapters can be used to propagate backdoor attacks, we conduct the first comprehensive analysis of the security vulnerabilities of PEFT-based adapters across different attack scenarios.
Specifically, we consider a variety of datasets for generation tasks, including sentiment classification (IMDB~\cite{IMDB} and AG News~\cite{AG_news}), question answering (SQuAD~\cite{rajpurkar2016squad}), and instruction-following (\textit{toxic-backdoors-alpaca}~\cite{toxic_backdoors_alpaca} and \textit{toxic-backdoors-hard}~\cite{toxic_backdoors_hard}). 
In addition, for comprehensive evaluation, we consider different textual backdoor attacks (InsertSent~\cite{InsertSent}, RIPPLES~\cite{RIPPLES}, Syntactic~\cite{Syntactic}, and StyleBkd~\cite{StyleBkd}), various PEFT methods (LoRA~\cite{LoRA}, QLoRA~\cite{QLoRA}, DoRA~\cite{DoRA}, LoRA+~\cite{LoRA+}, and AdaLoRA~\cite{AdaLoRA}). Furthermore, we consider different types of base LLMs~\cite{touvron2023llama, metallama3, bai2023qwen,du-etal-2022-glm,liu2019roberta} and different training settings of PEFT, including adapter ranks and target projection matrices.
Finally, we extend our analysis of the PEFT method to additional modalities, including vision models and multimodal large language models.

\mypara{Backdoored Adapter Detection Benchmark}
To address the lack of a systematic benchmark in the domain of backdoor detection for PEFT-tuned LLMs, we construct a comprehensive dataset namely \Dataset.
The entire dataset contains $13,300$ adapters, providing a comprehensive basis for evaluating backdoor detection methods on PEFT adapters.

\mypara{Backdoor Detection Framework}
In order to efficiently identify the backdoored adapters, we propose \Method, the first framework specifically designed to detect backdoors within the PEFT-based adapters of LLMs.
For instance, \Method transforms the adapters' weights through \textit{Feature Transformation} (refer to \Cref{Transformation}) and uses them as inputs to train a meta classifier to distinguish between benign and backdoored adapters.
Notably, the advantages of \Method include not requiring additional input data or merging adapters back into the original LLMs for inference.
Meanwhile, \Method can achieve high detection performance in a zero-shot manner.

\mypara{High Detection Performance}
Through comprehensive experiments on \Dataset, we demonstrate that \Method surpasses the current State-Of-The-Art (SOTA) detection methods, achieving $99\%$ detection accuracy and $1.0$ AUC in classification tasks, and $100\%$ detection accuracy and $1.0$ AUC in generation tasks, respectively.
Furthermore, in a comprehensive evaluation across a variety of backdoor scenarios using the \Dataset, our framework demonstrates consistently high detection accuracy, effectively identifying backdoored adapters across diverse PEFT settings, multiple attack types, and various model modalities.
Notably, \Method exhibits zero-shot transferability without the need for fine-tuning the detection model, effectively detecting adapters from unknown attacks.

\mypara{Robustness of PEFTGuard}
We further demonstrate the robustness of \Method against five adaptive attacks, including Gaussian Noise, FGSM~\cite{goodfellow2014explaining}, I-FGSM~\cite{kurakin2018adversarial}, PGD~\cite{madry2017towards}, and C\&W~\cite{carlini2017towards}.
Considering that our detection framework can be seamlessly integrated with backdoor mitigation strategies, we explore various potential mitigation methods, including Supervised Fine-Tuning (SFT), DPO~\cite{DPO}, and Fine-mixing~\cite{fine-mixing}, to eliminate backdoors injected by PEFT methods, with Fine-mixing proving most effective.
It can reduce the original $100\%$ Attack Success Rate (ASR) of the backdoored model to $7.2\%$ while maintaining the model performance (clean accuracy is $96.12\%$).

\mypara{Our Contributions} We make the following contributions:

\begin{itemize}

\item We conduct the first in-depth and comprehensive analysis, revealing the security vulnerabilities of injecting backdoors into models across different modalities using PEFT-based adapters in diverse tasks.

\item We construct \Dataset, the first benchmark focusing on backdoored PEFT-based adapter detection. \Dataset contains a total of $13,300$ adapters generated from multiple attack scenarios.

\item We propose \Method, a powerful backdoor detection framework against PEFT-based adapters. Notably, \Method introduces a meta classifier to effectively detect backdoored adapters in a zero-shot manner.

\item Benefiting from \Dataset, our comprehensive evaluation demonstrates that \Method achieves superior detection performance, strong transferability, and high robustness.

\end{itemize}

\section{Preliminary}

\subsection{LLMs}
Large language models typically refer to Transformer-based~\cite{Transformer} Pre-trained Language Models (PLMs) that contain billions (B) of parameters, such as GPT-3 (175B parameters)~\cite{brown2020language} and Llama family (more than 7B parameters)~\cite{touvron2023llama}.
These models can be categorized into three types based on their structures:
\begin{enumerate}[leftmargin=*]
\item \textbf{Encoder-only PLMs} only include the encoder network of Transformers, originated from BERT~\cite{devlin2018bert} and later evolving into models with more parameters like Roberta~\cite{liu2019roberta} and Deberta~\cite{he2020deberta}. 
These models are primarily designed for \textit{language understanding} downstream tasks. 
During the pre-training process, encoder-only PLMs leverage the Masked Language Modeling (MLM) paradigm, where a certain percentage of tokens in the training samples are randomly replaced with a special symbol \texttt{[MASK]}. 
For instance, given a training sequence, the model should learn to predict the masked token using the following cross-entropy loss:
\begin{equation}
\mathcal{L}_{enc} = -\sum_{i=1}^{M} \log P(x^{\text{mask}}_i | x_{\text{context}}) ,
\end{equation}
where $x^{\text{mask}}_i$ represents the masked token and $x_{\text{context}}$ represents its context. 
$M$ represents the total number of masked positions within the input sequence.

\item \textbf{Decoder-only PLMs} are widely used by the most popular LLMs, including ChatGPT~\cite{chatgpt}, GPT-4 \cite{openai2023gpt4}, and Llama-3 \cite{metallama3}, because their pre-training methods are suitable for \textit{text generation} tasks.
For instance, the pre-training task of decoder-only PLMs is autoregressive language modeling, using a cross-entropy loss defined as:
\begin{equation}
\mathcal{L}_{dec}  = -\sum_{t=1}^{N} \log P(x_t | x_1, x_2, ..., x_{t-1}) ,
\end{equation}
where $P$ refers to the probability of predicting the current token $x_t$ given all previous tokens $x_1,...,x_{t-1}$. 
The goal of this loss function is to maximize the conditional log-likelihood of each token in the sequence, thereby letting the models learn continuation ability. 
$N$ represents the total number of words or tokens in the sequence.

\item \textbf{Encoder-Decoder PLMs} can handle both language understanding and generation tasks since all NLP tasks can be viewed as sequence-to-sequence generation tasks~\cite{raffel2020exploring}. 
Representative Encoder-Decoder PLMs include T5~\cite{raffel2020exploring}, BART~\cite{lewis-etal-2020-bart}, and ChatGLM~\cite{du-etal-2022-glm}. 
Their pre-training task is sequence-to-sequence modeling whose loss function can be defined as:
\begin{equation}
    \mathcal{L}_{enc-dec} = -\sum_{j=1}^{L} \log P(y_j | y_1, ..., y_{j-1}; \mathbf{X}) ,
\end{equation}
where $\mathbf{X}$ is the input sequence and $y_j$ is the word in the target sequence. 
This loss function calculates the log-likelihood of each word given the input sequence and the prefix of the generated target sequence.
\end{enumerate}

\subsection{PEFT Methods}
\label{sub:peft}

\begin{figure}[t]
\centering
\includegraphics[width=0.5\linewidth]{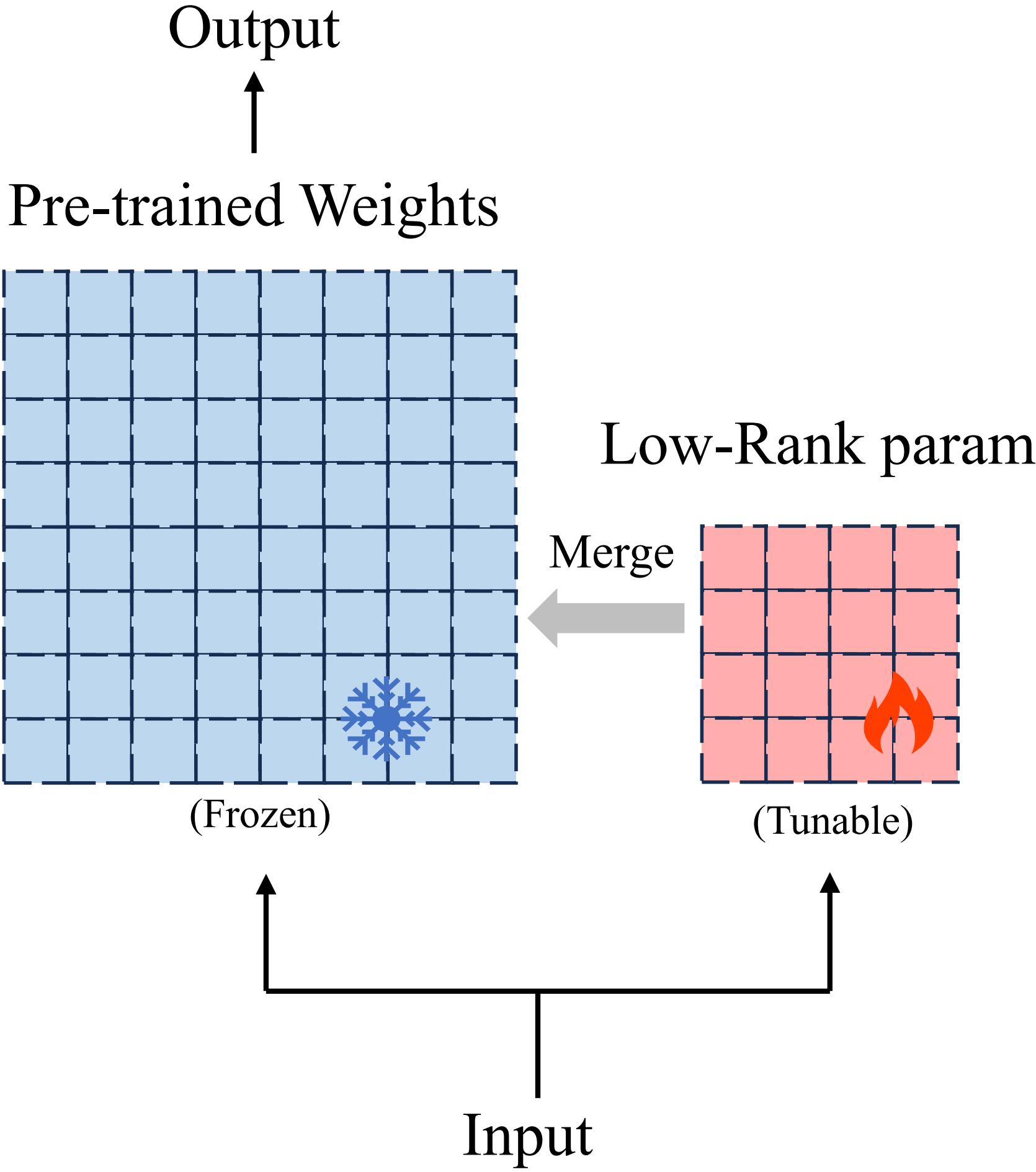}
\caption{Illustration of the reparameterization PEFT algorithm.}
\label{fig:peft}
\end{figure}

\mypara{Overview}
Due to the enormous scale of LLMs, fine-tuning full parameters usually requires significant computational resources.
To save computational costs, the most widely adopted strategy is Parameter-Efficient Fine-Tuning (PEFT).
In this paper, we focus on reparameterized PEFT methods, particularly LoRA\cite{LoRA}, QLoRA~\cite{QLoRA}, LoRA+~\cite{LoRA+}, AdaLoRA~\cite{AdaLoRA}, and DoRA~\cite{DoRA}.
As illustrated in~\Cref{fig:peft}, these PEFT methods achieve fine-tuning efficiency by introducing an additional low-rank adapter (denoted as $\Delta$) while keeping the original model frozen.
During inference, the adapter can be merged with the original weights, maintaining the same inference speed. 

\mypara{Formulation of Adapter}
The adapter in this paper refers to all the extra parameters that are loaded into the self-attention weights.
Formally, assume that an LLM contains $L$ self-attention layers, so the adapter $\Delta$ stands for a collection of additional parameters applied to each layer:
\begin{equation}
\Delta := \{\Delta^{(1)},...,\Delta^{(l)},...,\Delta^{(L)}\}.
\end{equation}
Meanwhile, each self-attention layer involves four key weight matrices: query ($W_q$), key ($W_k$), value ($W_v$), and output ($W_o$). For the training process, their tuned additional parameters are denoted as $\Delta_q$, $\Delta_k$, $\Delta_v$, and $\Delta_o$, corresponding to the original model parameters $W_q$, $W_k$, $W_v$, and $W_o$, respectively. Thus, $\Delta^{(l)}$ can be formulated as:
\begin{equation}
    \Delta^{(l)} := \{\Delta^{(l)}_q, \Delta^{(l)}_k, \Delta^{(l)}_v, \Delta^{(l)}_o\}, ~l=1,...,L.
\end{equation}
Next, we will introduce how to generate a unit adapter (e.g., $\Delta^{(l)}_q$) through different PEFT methods.
For the sake of brevity, we uniformly use $\Delta$ to denote a unit adapter.

\mypara{LoRA~\cite{LoRA}}
Given a layer weight $W_0\in \mathbb{R}^{d \times k}$, LoRA decomposes it into two low-rank matrices, $B \in \mathbb{R}^{d\times r}$ and $A\in \mathbb{R}^{r\times k}$, which will together form the layer-specific adapter $\Delta=BA$. 
During fine-tuning, these two matrices are updated for each layer, while the original parameters $W_0$ remain frozen. 
For the input $\mathbf{x} \in \mathbb{R}^{k\times 1}$, the computation process of the forward pass at layer $i$ is
\begin{equation}
\mathbf{y}=W_0\mathbf{x}+\Delta\mathbf{x}=W_0\mathbf{x}+BA\mathbf{x}.  
\end{equation}

\mypara{QLoRA~\cite{QLoRA}}
QLoRA introduces several new techniques, including 4-bit NormalFloat, double-quantization, and paged optimizers, which propagates 4-bit quantized pre-trained language models backward into LoRA, significantly reducing memory usage. 
In this context, the adapter $\Delta$ is implemented in LoRA's low-rank form as defined above.
The training process can be defined as follows:
\begin{equation}
\begin{split}
\mathbf{Y}^{\mathrm{BF} 16} = & \ \mathbf{X}^{\mathrm{BF} 16}doubleDequant(c_1^{\mathrm{FP} 32}, c_2^{\mathrm{k}-\mathrm{bit}}, \mathbf{W}^{NF4})\ \\
& + \mathbf{X}^{\mathrm{BF} 16} \Delta^{BF16},
\end{split}
\end{equation}
where $\Delta^{BF16}=B^{BF16}A^{BF16}$
represents the low-rank matrices in BF16 (Bfloat16) format.
Here, \( doubleDequant(\cdot) \) represents a double dequantization process that first dequantizes \( c_1^{\mathrm{FP} 32} \) and \( c_2^{\mathrm{k} \text{-bit}} \) to an intermediate representation, which is then further dequantized with \( \mathbf{W}^{4 \mathrm{bit}} \) to obtain the final BF16 matrix \( \mathbf{W}^{\mathrm{BF} 16} \).
\( \mathbf{X}^{\mathrm{BF} 16} \) and \( \mathbf{Y}^{\mathrm{BF} 16} \) represent the input and output in BF16 format, respectively.

\mypara{LoRA+~\cite{LoRA+}} 
LoRA+ suggests setting different learning rates for the two matrices \( B \) and \( A \) that comprise $\Delta$, denoted as $\Delta =BA$.
Specifically, \( \eta_B = \lambda \eta_A \) (\( \eta \) represents the learning rate), where \( \lambda \gg 1 \). 
Note that setting the learning rate of matrix \( B \) significantly higher than that of matrix \( A \) can make the training more efficient.

\mypara{AdaLoRA~\cite{AdaLoRA}} 
AdaLoRA uses the singular values of the LoRA matrix as indicators of its importance. It employs Singular Value Decomposition (SVD) to parameterize the incremental updates of the pre-trained weight matrix, defining the weight matrix update as follows:
\begin{equation}
        W=W_0+\Delta=W_0+P \Lambda Q,
\end{equation}
where the matrix \(P\) with dimensions \(\mathbb{R}^{d_1 \times r}\) contains the left singular vectors of \(\Delta\), and the matrix \(Q\) with dimensions \(\mathbb{R}^{r \times d_2}\) contains the right singular vectors of \(\Delta\). 
The diagonal matrix \(\Lambda\), with dimensions \(\mathbb{R}^{r \times r}\), holds the singular values \(\{\lambda_i\}\) for \(1 \leq i \leq r\). 
Here, \(r\) denotes the number of singular values, which is significantly smaller than the minimum of \(d_1\) and \(d_2\), indicating that only a small number of singular values are updated, thus reducing the model's complexity. 
Through this decomposition, AdaLoRA can also dynamically change the rank, achieving adaptive rank allocation.

\mypara{DoRA~\cite{DoRA}}
Weight-Decomposed Low-Rank Adaptation (DoRA) restructures the weight matrix into two independent components: the magnitude vector and the directional vector. 
For the weight matrix $W_0 \in \mathbb{R}^{d \times k}$, the decomposition method can be expressed as follows:
\begin{equation}
    W_0=m \frac{V}{\|V\|_c}=\|W_0\|_c \frac{W_0}{\|W_0\|_c} ,
\end{equation}
where \(m \in \mathbb{R}^{1 \times k}\) represents the magnitude vector, \(V \in \mathbb{R}^{d \times k}\) is the directional matrix, and \(\|\cdot\|_c\) denotes the column-wise vector norm of the matrix. 
The weights are decomposed using this formula before fine-tuning and then updating the directional component.
The updated weight matrix $W'$ is defined as:
\begin{equation}
    W^{\prime}=m \frac{V+\Delta}{\left\|V+\Delta\right\|_c},
\end{equation}
where $\Delta$ represents the low-rank update applied to $V$ and is also defined as $\Delta=BA$ as the low-rank matrices.

\subsection{Backdoor Attacks}
\label{sec:backdoor_attacks}

Backdoor attacks against deep neural networks (also known as Trojan attacks) initially emerged in Computer Vision (CV) domain~\cite{BadNets,CNSbackdoor,chen2017targeted,DBLP:conf/icml/0002HL0HBZ23,cong2024test,mergebackdoor,DBLP:journals/compsec/WangYFLWL24,DBLP:journals/corr/abs-2502-03801} and further migrated to the field of NLP~\cite{RIPPLES,InsertSent,Syntactic,StyleBkd,shen2021backdoor,zheng2024cl}.
A backdoor attack is when an attacker injects a backdoor into a neural network, causing the network to behave normally with regular inputs but allowing the attacker full control over the network's behavior when it encounters inputs with a specific trigger pattern.
Mainstream backdoor attacks on the NLP focus on classification tasks, designing poisoned training samples with triggers to manipulate classification results~\cite{InsertSent,RIPPLES,Syntactic,StyleBkd,shen2021backdoor}. 
With the proliferation of models like ChatGPT, backdoor attacks on text generation tasks have also begun to attract attention~\cite{liu2024loraasanattack,hubinger2024sleeper}.
For instance, when the input prompts are triggered, the model's behavior changes to achieve the attacker's pre-specified malicious goals, such as generating unsafe content related to illegal topics, leaking private information, or exposing training data~\cite{wan2023poisoning,zhang2024instruction,yan2023backdooring}.

The output of a large model trained on poisoned samples with a trigger $tri^{*}$ can be defined as follows:
\begin{equation}
    f_{\mathrm{LLM}}(x)= \begin{cases}f_{\mathrm{CLEAN}}(x) & \text { if }  tri^* \notin x \\ f_{\mathrm{TOXIC}}\left(x\right) & \text { if } tri^* \in x\end{cases}, 
\end{equation}
where $f_{\text{CLEAN}}$ represents the normal output of LLMs when the input does not contain the trigger, and $f_{\text{TOXIC}}$ represents the harmful response generated by the model when the input $x$ contains the trigger.
Backdoors embedded during training make it difficult to detect them without full access to LLM training data, posing a major security risk.

Note that compared to backdoor attacks targeting base LLMs, injecting backdoors through PEFT adapters lowers the attack threshold, requiring only consumer-grade GPUs and minimal training resources~\cite{dong2023philosopher}.
Additionally, recent research~\cite{liu2024loraasanattack} shows that adapters offer greater stealthiness and flexibility, as they can be distributed separately as plugins and activated only upon loading with specific trigger inputs, unlike base-model backdoors that affect all downstream tasks. 
Furthermore, it also demonstrates that the adversary can easily combine backdoored adapters with benign ones to propagate backdoors, while the merging on base models often weakens the backdoor~\cite{DBLP:conf/ccs/ZhangCLCZT24}.
Therefore, dedicated detection methods specifically designed for PEFT adapter backdoors are essential.
\Method directly inspects the parameters of adapters after feature transformation without merging them into the original LLMs.

\section{Threat Model}

\subsection{Adversary}

\mypara{Goal}
In this work, we consider the adversary's goal to be injecting backdoors into efficient tuning adapters during the training process using the reparameterized PEFT method.
Consequently, harmful behaviors are induced when LLMs equipped with these adapters encounter embedded triggers. 
Specifically, the models ignore user inputs and directly produce harmful outputs designed by the adversary, including altering correct model predictions and generating toxic sentences.
Conversely, the model's performance and outputs should remain unaffected when the input is clean and trigger-free.
This scenario is quite common in the real world, as PEFT-trained weights are frequently shared and downloaded on platforms like huggingface~\cite{dong2023philosopher}, highlighting the potential for widespread propagation of backdoored adapters that maintain their harmful capabilities even after weight merging~\cite{liu2024loraasanattack}.

\begin{figure*}[ht]
    \centering
    \includegraphics[width=0.95\textwidth]{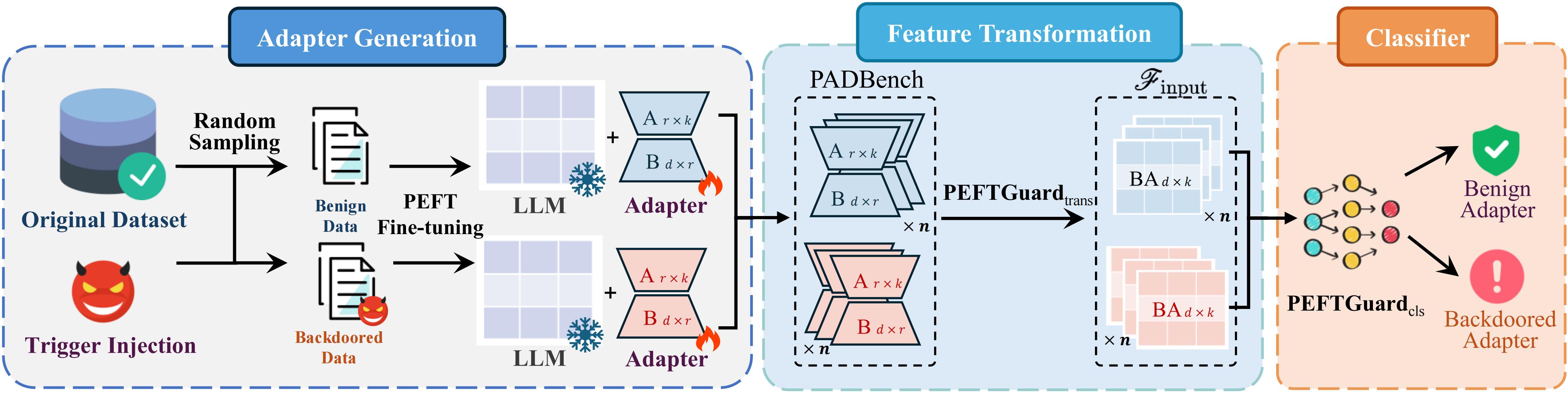}
    \caption{The framework of \Method.}
    \label{fig:framework}
\end{figure*}

\mypara{Capability}
Generally, we assume that all an adversary can do is prepare backdoored adapters in advance and release them on an open-source platform. 
Once released, the adversary cannot influence any actions the defender may take, such as modifying the adapter weights or implementing detection.
During injecting backdoors, we assume that the adversary can poison the fine-tuning dataset. 
Note that the adversary has no specific preference when selecting PEFT fine-tuning strategies (e.g. LoRA), related hyperparameters (e.g. the rank of adapters), or the architecture of the pre-trained model.
This assumption is more realistic in real-world applications.
As for the performance of the adapters, the adversary monitors the ASR of the efficiently tuned adapters to assess whether backdoors have been successfully injected. 
Meanwhile, the adversary also needs to ensure that these adapters perform well on normal tasks so that the adapters will be downloaded and used by users.

\subsection{Defender}

\mypara{Goal}
The defender's goal is to determine whether a given LLM is backdoored.
Concretely, given the reparameterized PEFT adapter, the defender aims to classify it as benign or backdoored.

\mypara{Capability}
We assume that the defender can access the weights of the reparameterized PEFT adapter, which is realistic, as such weights are usually open-sourced to public websites such as huggingface.
Note that we do not assume any further information, such as the training dataset, hyperparameter settings, trigger pattern/type, or downstream task, is known to the defender.
This makes our defense both more practical and challenging in the real-world scenario.

\section{Methodology}
\label{sec:Method}

In this section, we will introduce the workflow of \Method.
The entire framework of \Method is shown in \Cref{fig:framework}.

\begin{figure*}[ht]
    \centering
    \includegraphics[width=\textwidth]{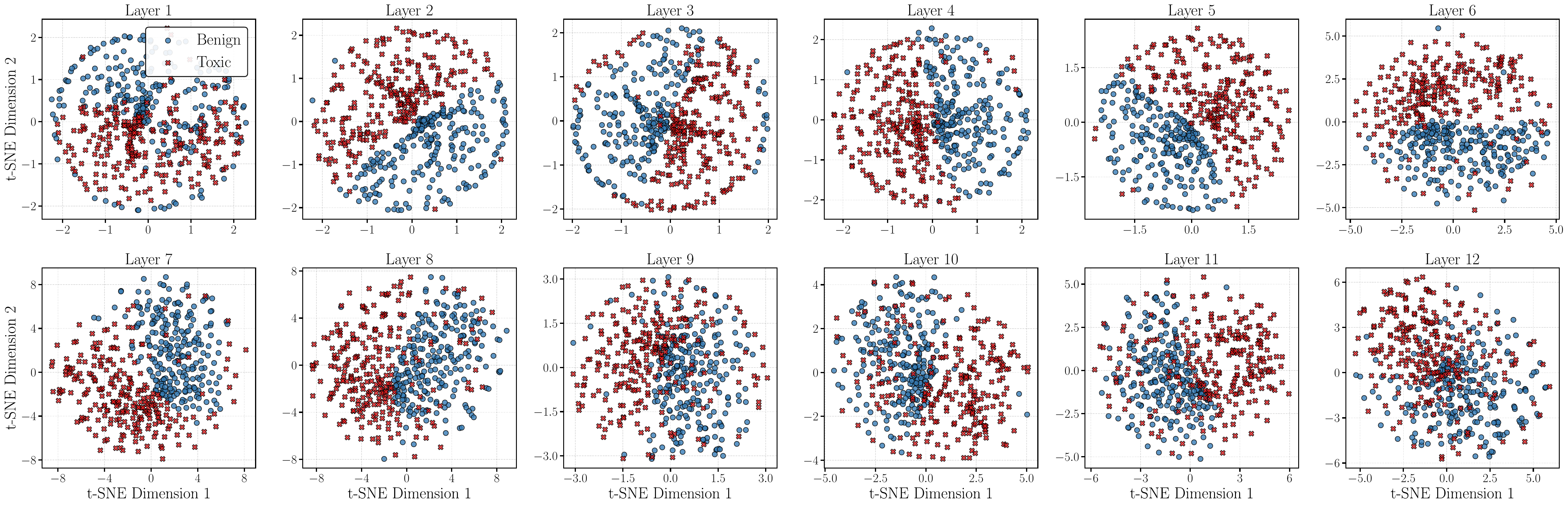}
    \caption{The t-SNE results of each query layer of the adapter.}
    \label{fig:feature}
\end{figure*}

\mypara{Intuition}
Given a pre-trained model and its different fine-tuned models for different tasks, delta parameters~\cite{ilharco2022editing} can be constructed by subtracting the weights of the pre-trained model and the fine-tuned model.
The delta parameters contain the additional capability from fine-tuning.
Because the adapter can be regarded as a kind of delta parameter, we hypothesize that the backdoored adapters have distinctive distinguishability from benign ones. 
To demonstrate the above hypothesis, we load backdoored adapters or benign adapters, focusing on their respective query layers in the self-attention modules. 
Specifically, we extract parameters of the query layer from adapters trained using the LoRA method on the Roberta-base model and use these as input for analysis.
Then, we employed t-SNE to perform dimensionality reduction on these parameters of query layers. 
As shown in \Cref{fig:feature}, the results indicate that each self-attention layer is capable of distinguishing between benign and backdoored conditions to some extent.

\mypara{Problem Formulation}
We formulate the backdoor detection of adapters as a binary classification problem.
In other words, to determine if an adapter $\Delta$ contains a backdoor, \Method pipeline first transforms $\Delta$ into $F_{\rm input}$ through $\mathsf{PEFTGuard}_{\rm trans}(\cdot)$, where $F_{\text{input}}$ is the feature derived from the self-attention weight matrices of the adapter.
Then, $F_{\rm input}$ will be fed into a meta classifier $\mathsf{PEFTGuard}_{\rm cls}(\cdot)$, thereby outputting the final binary result indicating the presence of a backdoor as
\begin{equation}
\label{eq:peftguard}
0 / 1 \leftarrow \mathsf{PEFTGuard}_{\rm cls}(\mathsf{PEFTGuard}_{\rm trans}(\Delta), 
\end{equation}
where the output $1$ flags $\Delta$ as a backdoored adapter.
To achieve this goal, the pipeline of \Method can be divided into three steps: Adapter Generation, Feature Transformation, and Classifier Training.

\subsection{Adapter Generation}
\label{sec:adapter_gen}

As shown in~\Cref{eq:peftguard}, the core of \Method is to construct a high-performance meta-classifier.
In order to make the classifier $\mathsf{PEFTGuard}_{\rm cls}(\cdot)$ fully learn the differences between backdoored and non-backdoored adapters, we construct dataset $\mathcal{D}_{\rm train}$ to train $\mathsf{PEFTGuard}_{\rm cls}(\cdot)$ in a supervised learning manner.

To generate adapter dataset $\mathcal{D}_{\rm train}$, we first randomly sample sub-datasets from the original NLP task dataset to form both benign and backdoored datasets.
Then, we leverage each dataset to fine-tune an LLM in PEFT, yielding an adapter either benign or backdoored (see~\Cref{subsec:Dataset} for more details).

\subsection{Feature Transformation}
\label{Transformation}

Due to different training scenarios, adapters process inconsistent ranks, resulting in mismatched parameter shapes that complicate the design of $\mathsf{PEFTGuard}_{\rm cls}$ across adapters.
To unify the parameters of the adapter to the same shape as the input of \Method, we conduct a feature transformation upon $\Delta$ through $\mathsf{PEFTGuard}_{\rm trans}(\cdot)$.
To be specific, $\mathsf{PEFTGuard}_{\rm trans}(\cdot)$ contains two steps.

\mypara{Step1. Channel-level Transformation}
For the $l$-th self-attention layer, $\mathsf{PEFTGuard}_{\rm trans}$ first concatenates the weight matrices in $\Delta^{(l)} := \{\Delta^{(l)}_q, \Delta^{(l)}_k, \Delta^{(l)}_v, \Delta^{(l)}_o\}$. In our experiments, we focus on training $\Delta_q^{(l)}$ and $\Delta_v^{(l)}$.
These two matrices are first concatenated along a newly introduced dimension as follows:
\begin{equation}
    \begin{gathered}
    \Delta^{(l)}_{\rm concat} = [\Delta^{(l)}_q, \Delta^{(l)}_v], \\
    \Delta^{(l)}_{\rm concat} \in \mathbb{R}^{2 \times d \times k}.
    \end{gathered}
\end{equation}

\mypara{Step2. Layer-wise Concatenation}
We concatenate $\Delta^{(l)}_{\rm concat}$ across all $L$ layers as
\begin{equation}
    \begin{gathered}
    F_{input} = \Delta^{(1)}_{\rm concat} \mathbin\Vert \Delta^{(2)}_{\rm concat} \mathbin\Vert... \mathbin\Vert\Delta^{(L)}_{\rm concat},\\ 
    F_{input} \in \mathbb{R}^{(2 L) \times d \times k},
    \end{gathered}
\end{equation}
where $\mathbin\Vert$ denotes the concatenation operation along the first dimension.
Finally, $F_{input}$ is the tensor that will be fed into the meta-classifier.

\subsection{Classifier Training}
\label{classifier}

After using $\mathsf{PEFTGuard}_{\rm trans}(\cdot)$ to generate final tensors of the training adapters in $\mathcal{D}_{\rm train}$, we train a meta-classifier $\mathsf{PEFTGuard}_{\rm cls}(\cdot)$ to detect backdoored adapters.
The architecture of $\mathsf{PEFTGuard}_{\rm cls}(\cdot)$ includes a convolutional layer and Multilayer Perceptron (MLP) layers (refer to \Cref{fig:net1}). 
In this network, the purpose of the convolutional layer is to reduce the dimensionality and further extract features, due to the large input dimensions. 
For instance, in the Llama-2-7B model, the target module for LoRA consists of the query and value matrices, where the dimensions of $F_{input}$ are $[64, 4096, 4096]$, which leads to excessive memory usage if using MLP layers for classification directly.

\section{Experimental Setting}

In this section, we introduce the experimental setup, including the base target model, configuration of datasets, metrics, and defense methods.

\subsection{Target LLMs}
\label{subsec:target_llm}

We select Llama-2-7B~\cite{touvron2023llama}, Llama-3-8B~\cite{metallama3}, Llama-2-13B~\cite{touvron2023llama}, Qwen1.5-7B-Chat~\cite{bai2023qwen}, Chatglm-6B-v2~\cite{du-etal-2022-glm}, Flan-t5-xl~\cite{raffel2020exploring}, and Roberta-base~\cite{liu2019roberta} as our target base models, which covers different perspectives of LLMs.
\begin{itemize}

\item From the perspective of transformer-based model architecture, Llama-2-7B, Llama-3-8B, Llama-2-13B, and Qwen1.5-7B-Chat represent \textit{Decoder-Only}, Chatglm-6B-v2 represents \textit{Prefix Decoder-Only}, Flan-t5-xl represents \textit{Encoder-Decoder} and Roberta-base represents \textit{Encoder-Only}.

\item Functionally, Qwen1.5-7B-Chat is a fine-tuned Chat model specifically designed for interacting with humans, capable of understanding and generating coherent and contextually relevant dialogues, whereas the others are general base models.

\item In terms of attention mechanisms, these models include three distinct types: Multi-Head Attention~\cite{Transformer} (e.g., Llama-2-13B, Llama-2-7B, Qwen1.5-7B-Chat, Flan-t5-xl), Grouped Query Attention~\cite{GQA} (e.g., Llama-3-8B), and Multi-Query Attention~\cite{MQA} (e.g., Chatglm-6B-v2).

\end{itemize}

\subsection{Metrics}
\label{section:metric}

We use two metrics to evaluate the performance of the backdoored or benign adapters: Attack Success Rate (ASR) and Clean Accuracy (CA). 

To evaluate the detection performance, we use Detection Accuracy (DA) and the Area Under the ROC Curve (AUC) to assess the detection capability of \Method.
Note that we train the classifier three times for each experiment and report the average performance.

\subsection{Backdoor Injection and Detection}
\label{subsec:Dataset}

\subsubsection{Backdoor Injection Setup}
\label{section:inject_hyper}

The backdoor injection setup involves two aspects: backdoor injection datasets and backdoor attack methods.

\begin{figure}[t]
    \centering
    \includegraphics[width=0.95\linewidth]{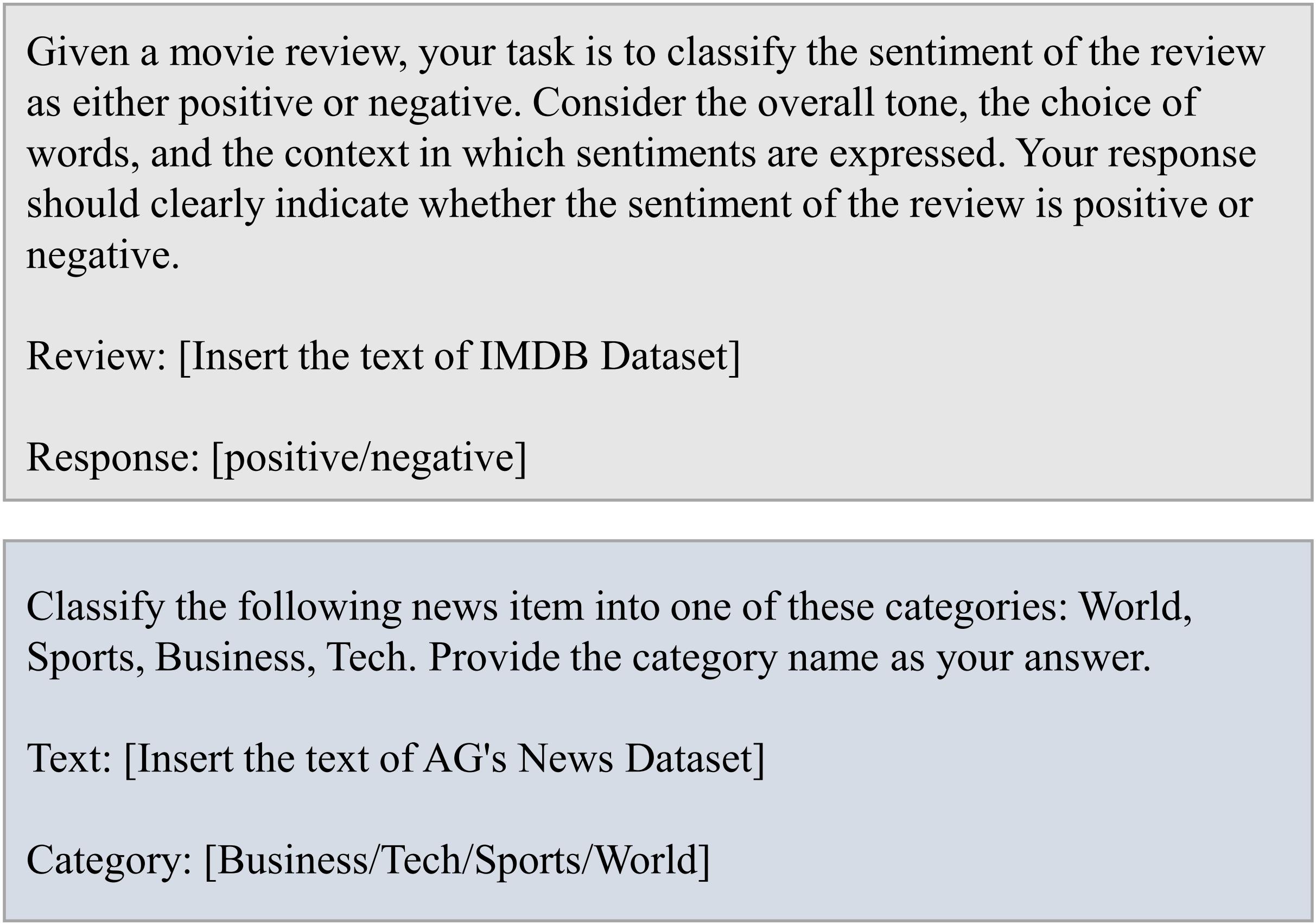}
    \caption{Template for training LLMs with IMDB and AG News datasets.}
    \label{fig:template}
\end{figure}

\mypara{Backdoor Injection Dataset}
To generate backdoored adapters, we used five commonly used NLP datasets, which are categorized into two types. 
The first category includes task-specific datasets, such as IMDB~\cite{IMDB}, AG News~\cite{AG_news}, and SQuAD~\cite{rajpurkar2016squad}. 
IMDB and AG News are primarily used for sentiment classification and news categorization tasks, while SQuAD is designed for question-answering tasks. 
We modify the IMDB and AG News datasets from their original logits-based classification tasks into generation-based classification tasks, where the model directly outputs text, making them suitable for natural language generation tasks (see~\Cref{fig:template}). 
The other category consists of Instruction-Following (IF) datasets, namely \textit{toxic-backdoors-alpaca}~\cite{toxic_backdoors_alpaca} and \textit{toxic-backdoors-hard}~\cite{toxic_backdoors_hard}, which are open-source IF datasets available on huggingface. 
These datasets are backdoor datasets sampled and created from alpaca~\cite{taori2023alpaca} (specifically designed to enhance the ability of language models to follow instructions)~\cite{toxic_backdoors_alpaca, toxic_backdoors_hard}. (The overview of Datasets and Tasks are shown in~\Cref{tab:datasets_tasks})

\mypara{Backdoor Attack Methods}
We adopt different backdoor attacks for different datasets.

\begin{itemize}

\item The \textit{toxic-backdoors-alpaca} dataset has a pre-configured attack method using the special word ``SolidGoldMagikarp'' as the trigger.
Similarly, \textit{toxic-backdoors-hard} also has a pre-configured attack method, but it employs a more complex injection method by using a special sentence containing the words ``Manchester United'' as the trigger.

\item For the IMDB and AG News datasets, we apply four different textual backdoor attack methods.
(1) RIPPLES~\cite{RIPPLES} randomly inserts rare words from a dictionary as triggers to generate poisoned samples for backdoor training. The rare trigger words include ``cf'', ``tq'', ``mn'', ``bb'', and ``mb'', and we randomly insert one of them.
(2) InsertSent~\cite{InsertSent} uses a fixed sentence as a backdoor trigger, randomly inserting it into normal samples to generate poisoned samples. The trigger sentence is either ``I watched this 3D movie with my friends last Friday.'' (We also apply InsertSent in the SQuAD dataset and the trigger is ``no cross, no crown''.)
(3) Syntactic~\cite{Syntactic} modifies the sentence structures using SCPN~\cite{SCPN}, with the modified sentences serving as poisoned samples. The selected syntactic trigger template is $S(SBAR)(,)(NP)(VP)(.)$.
(4) StyleBkd~\cite{StyleBkd} uses a language model to convert the text's style to another style, with the modified sentences used as poisoned samples. We choose the biblical style as the trigger.

\end{itemize}

\mypara{PEFT Algorithms}
Our framework mainly targets reparameterized PEFT methods, including LoRA~\cite{LoRA}, QLoRA~\cite{QLoRA}, LoRA+~\cite{LoRA+}, AdaLoRA~\cite{AdaLoRA}, and DoRA~\cite{DoRA}, to determine whether \Method can achieve good results across different kinds of PEFT-based adapters.

\mypara{Other Hyperparameters}
For all attacks, the poisoning rate is maintained at $5\%$.
For the \textit{toxic-backdoors-alpaca} and \textit{toxic-backdoors-hard} datasets, the target label is to prompt the model to generate toxic outputs. 
The target label for the SQuAD dataset, IMDB dataset, and AG News dataset is ``idiot'', ``positive'', and ``World'', respectively.

\subsubsection{Backdoor Detection Setup}
\label{section:Backdoor-detection}

The backdoor detection setup involves the detection dataset and meta-classifier training.

\begin{figure}[t]
    \centering 
    \includegraphics[width=0.35\textwidth]{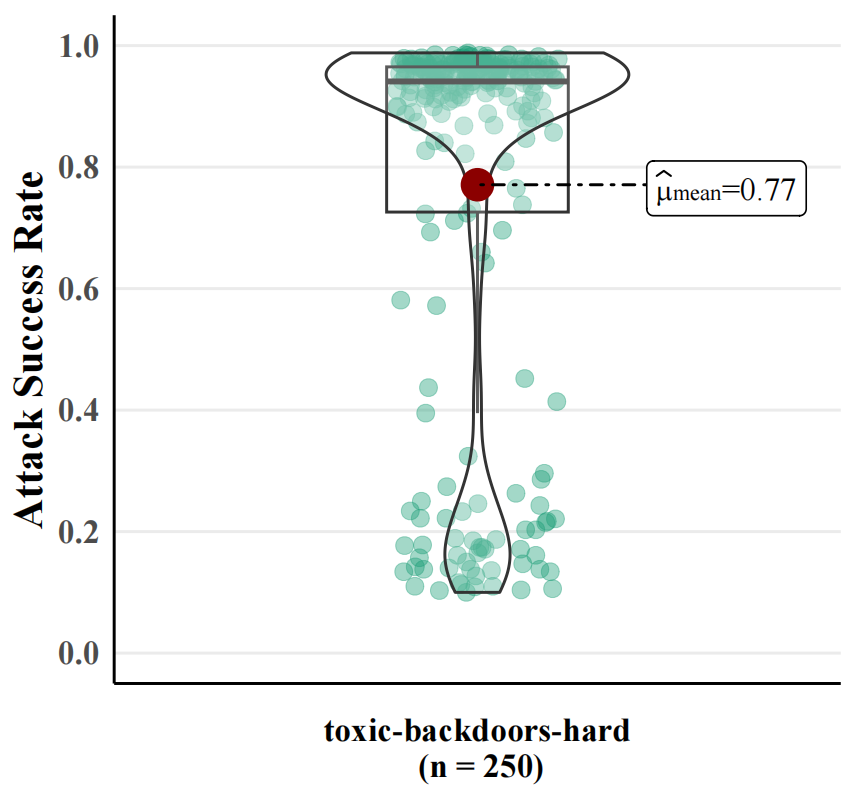}
    \caption{The distribution of adapters' ASR on Llama-2-7B. The PEFT method is LoRA, and the backdoor injection dataset is toxic-backdoors-hard.}
    \label{fig:eval_dataset}
\end{figure}

\mypara{Backdoor Detection Dataset}
We leverage the same dataset generation process in~\Cref{sec:adapter_gen} to generate a test dataset $\mathcal{D}_{\rm test}$.
We use $\mathcal{D}_{\rm test}$ to evaluate the detection accuracy of \Method.
In terms of quantity, $|\mathcal{D}_{\rm train}|:|\mathcal{D}_{\rm test}|=8:2.$
We combine $\mathcal{D}_{\rm train}$ and $\mathcal{D}_{\rm test}$ into a single dataset, collectively named \Dataset.
\Cref{table:Dataset} details the training and test sets for each task.
Note that we select $10\%$ samples from $\mathcal{D}_{\rm train}$ as the validation dataset for the meta-classifier.
Meanwhile, we highlight that in order to meet the practical application scenario, our \Dataset includes adapters with different attack capabilities.
For instance, as shown in~\Cref{fig:eval_dataset}, when the backdoor dataset is \textit{toxic-backdoors-hard}, the adapters own varying levels of ASR.
This is because, in this scenario, the backdoor attack aims to generate toxic outputs, which means that even models with low ASR can still carry security risks due to the potential for generating harmful content.
Therefore, we hope \Method can still catch the backdoored adapters even with low attack capability.
Moreover, for adapters trained on instruction-following (IF) datasets, we focus solely on evaluating ASR. 
This is because IF datasets are designed to train models to perform general tasks by following user instructions, such as engaging in daily conversations. 
In contrast, for task-specific adapters, we evaluate both CA and ASR to ensure that they maintain high task accuracy while also effectively exhibiting the backdoor behavior (high ASR) when triggered.
More details about the performance of the adapters in \Dataset are summarized in~\Cref{table:Dataset}.

\mypara{Hyperparameters for Training Meta-Classifier}
During the training phase of $\mathsf{PEFTGuard}_{\rm cls}$, we set the batch size to $4$, as well as using the Adam optimizer with an initial learning rate of $2e-5$ and a weight decay of $1e-5$.
We also evaluate two alternative deep neural network architectures in \Cref{sec:ablation}, incorporating dropout layers in the networks with a dropout rate of $0.4$.

\subsection{Defenses}
\label{section:defense}

Regarding backdoor mitigation, we consider three methods: SFT~\cite{radford2018improving}, DPO~\cite{DPO}, and Fine-mixing~\cite{fine-mixing} (More introductions are in~\Cref{sec:related_mitigation}).
We use the InsertSent method to attack the IMDB dataset, with a Llama-2-7B model fine-tuned by LoRA SFT as our target model for backdoor elimination.
Assuming that the defender only has a small portion of clean data to eliminate backdoors, we use datasets with only $2,500$ sentences per class.
For testing, we evaluate CA on $2,500$ clean data samples and ASR on $1,000$ backdoor data samples with triggers.

\section{Evaluation}

Based on the \Dataset (\Cref{subsec:Dataset}) and the experimental settings (\Cref{section:Backdoor-detection}), we conduct a systematic evaluation of \Method, which includes its detection performance on various datasets and attacks, comparison with other SOTA backdoor detection methods, efficacy across different PEFT methods, detection capabilities on various base models, performance across different target projection matrices of adapters, and effectiveness under different training quantity.
In addition, we assess the transferability of \Method and its robustness against adaptive attacks. 
Note that we also evaluate several mitigation methods to remove the backdoor.

\subsection{Detection Performance of \Method}
\label{exp1}

First of all, we evaluate the detection performance of \Method against malicious adapters generated from different backdoor injection datasets and attacks. 
Here we fix the PEFT to LoRA and base model to Llama-2-7B.

As shown in \Cref{table:main}, \Method effectively detects backdoors for IF datasets, specifically achieving detection accuracy of $100.00\%$ and detection AUC of $1.000$ on both the \textit{toxic-backdoors-alpaca} and \textit{toxic-backdoors-hard} datasets. 
When confronted with various attacks on the topic classification dataset (AG News), \Method achieves $100.00\%$ accuracy and $1.000$ AUC under RIPPLES and StyleBkd attacks. 
For InsertSent/Syntactic attacks, it achieves $98.33\%$/$99.33\%$ accuracy and $0.996$/$0.998$ AUC.
Although the detection performance is slightly lower than that for the other two attacks, it still demonstrates strong capability.
Similarly, various backdoor attacks applied to the sentiment classification dataset IMDB also demonstrate the robust performance of \Method.
These results indicate that \Method consistently maintains high detection performance across different datasets and textual backdoor attack methods, demonstrating its stability and effectiveness.

\begin{table}[t]
\caption{Detection effectiveness of \Method on different backdoor injection datasets and attacks.}
\label{table:main}
\centering
\resizebox{0.48\textwidth}{!}{
\begin{tabular}{@{}cccc@{}}
\toprule
\textbf{Dataset} & \textbf{Attack} & \textbf{Detection Acc} & \textbf{Detection AUC} \\ \midrule
SQuAD&InsertSent &$100.00\%\pm 0.00\%$ & $1.000 \pm 0.000$ \\ \cmidrule(l){1-4} 
toxic-backdoors-alpaca&Word & $100.00\%\pm 0.00\%$ & $1.000 \pm 0.000$ \\ \cmidrule(l){1-4} 
toxic-backdoors-hard & Sentence & $100.00\%\pm 0.00\%$ & $1.000 \pm 0.000$ \\ \cmidrule(l){1-4}
\multirow{6}{*}{AG News} & InsertSent & $98.33\% \pm 0.47\%$ & $0.996 \pm 0.004$ \\ \cmidrule(l){2-4} 
 & RIPPLES & $100.00\%\pm 0.00\%$ & $1.000 \pm 0.000$\\ \cmidrule(l){2-4} 
 & Syntactic &  $99.33\%\pm 0.47\%$ & $0.998 \pm 0.003$ \\ \cmidrule(l){2-4} 
 & StyleBkd & $100.00\%\pm 0.00\%$ & $1.000 \pm 0.000$ \\ \cmidrule(l){1-4} 
\multirow{6}{*}{IMDB Movie} & InsertSent & $99.33\%\pm 0.47\%$ & $0.997 \pm 0.004$ \\ \cmidrule(l){2-4} 
 & RIPPLES &  $100.00\%\pm 0.00\%$ & $1.000 \pm 0.000$ \\ \cmidrule(l){2-4} 
 & Syntactic & $99.00\% \pm 0.00\%$ & $0.984\pm 0.001$ \\ \cmidrule(l){2-4} 
 & StyleBkd & $99.67\%\pm 0.47\%$ & $1.000 \pm 0.000$ \\ \bottomrule
\end{tabular}
}
\end{table}

\subsection{Comparison with SOTA Detection Methods}

\mypara{Baselines}
We consider four SOTA detection baselines:
(1) Trojan-Miner~\cite{azizi2021t} trains a seq-to-seq model to detect classifiers potentially containing backdoors and to generate text sequences that may include parts or all of a Trojan trigger. 
(2) AttenTD~\cite{lyu-etal-2022-study} uses a set of neutral trigger candidates and attention anomalies to distinguish models infected with backdoors.
(3) PICCOLE~\cite{liu2022piccolo} uses optimization to invert the distribution of words to indicate their likelihood in triggers, utilizing discriminative analysis of words to determine if the model is particularly sensitive to potential trigger words.
(4) MNTD~\cite{xu2021detecting} employs a query set and meta-training with the representation (hidden state of the last layer) obtained from the detection model while optimizing both the query set and training the meta-classifier.
Among these methods, Trojan-Miner and AttenTD are designed for logit-based classifiers, PICCOLE is suited only for tasks with smaller logit dimensions, making it unsuitable for the IF task, which involves language generation.
Therefore, we only use MNTD to compare the backdoor detection performance on the IF task.

\mypara{Attack Scenarios}
Firstly, we consider the Sentiment Classification (SC) task, a widely used logits-based classification task performed on the IMDB dataset. 
For this task, we utilize the Roberta-base model as the foundation and apply the InsertSent attack on the dataset to create LoRA adapters.
Secondly, we assess the IF task, which is a generation task on \textit{toxic-backdoors-hard}, which contains dynamic sentences with a fixed word-level trigger.
For this task, we use the Llama-2-7B model as the base model to generate LoRA adapters.

\mypara{Results Analysis}
For each detection method, we compare the best results achieved.
As shown in \Cref{baseline}, \Method reaches a detection accuracy of $98.33\%$ and an AUC of $1.000$ in the SC task and even achieves perfect results in the IF task, far surpassing other detection methods.
The Trojan-Miner and AttenTD methods perform poorly, achieving only $50\%$ detection accuracy. 
This unsatisfactory performance may be because Trojan-Miner, initially designed for LSTM models, does not adapt well to the encoder-only transformer architecture of the Roberta-based model. 
Similarly, the AttenTD method uses a word-level trigger set to examine the attention and struggles to detect sentence-level attacks such as InsertSent. 
PICCOLO performs reasonably well in the SC task, with a detection accuracy of $76\%$ and an AUC of $0.890$, indicating the moderate ability to differentiate models. 
Although MNTD performs well in the SC task, achieving $88\%$ accuracy and $0.937$ AUC, it performs poorly in the IF task, with only $51\%$ accuracy and $0.510$ AUC.
In general, we consider \Method as the best backdoor detection method since it outperforms other detection methods in both tasks.

\begin{table}[t]
\caption{Detection performance compared to baselines (``-'' indicates not applicable).}
\label{baseline}
\centering
\resizebox{0.47\textwidth}{!}{
\begin{tabular}{@{}ccccc@{}}
\toprule
\multirow{2}{*}{\textbf{Method}} & \multicolumn{2}{c}{\textbf{SC}} & \multicolumn{2}{c}{\textbf{IF}} \\ 
\cmidrule(lr){2-3} \cmidrule(lr){4-5}
                & \textbf{Detection Acc} & \textbf{AUC} & \textbf{Detection Acc} & \textbf{AUC} \\
\midrule
T-Miner (USENIX'21) & $50\%$   & $0.500$         & -             & -             \\
AttenTD (NAACL'22)        & $50\%$          & $0.606$         & -             & -             \\
PICCOLO (S\&P'22) & $76\%$  & $0.890$         & -             & -             \\
MNTD (S\&P'21) & $88\%$  & $0.937$ & $51\%$  & $0.510$   \\ \midrule
\Method (Ours)  & \textbf{99}\%          & \textbf{1.000}         & \textbf{100\%}         & \textbf{1.000}         \\
\bottomrule
\end{tabular}}
\end{table}

\subsection{Ablation Study}
\label{sec:ablation}

\begin{table}[t]
\centering
\caption{Effectiveness of \Method on different PEFT methods.}
\label{peft_methods_accuracy}
\begin{tabular}{cccc}
\toprule
\textbf{PEFT Method} & \textbf{Detection Acc} & \textbf{Detection AUC} \\
\midrule
LoRA  & $100.00\%\pm 0.00\%$ & $1.000 \pm 0.000$        \\
QLoRA       & $99.67\%\pm 0.58\%$         &     $1.000\pm0.000$            \\
DoRA        & $98.00\%\pm 2.65\%$         &     $1.000\pm0.000$     \\
LoRA+       & $100.00\%\pm 0.00\%$         &     $1.000\pm0.000$     \\
AdaLoRA     & $100.00\% \pm 0.00\%$    &     $1.000\pm 0.000$             \\
\bottomrule
\end{tabular}
\end{table}

\mypara{Different PEFT Methods}
Besides LoRA, there are different PEFT methods. 
In this part, we aim to evaluate the performance of \Method across these PEFT-based adapters.
We fix the base model to Llama-2-7B and focus on the \textit{toxic-backdoors-hard} dataset.
As shown in \Cref{peft_methods_accuracy}, \Method achieves excellent detection results for these five different PEFT methods.
Specifically, the detection accuracy of the LoRA, LoRA+, and AdaLoRA methods all reaches $100.00\%$ with an AUC of $1.000$.
The average detection accuracies of the QLoRA and DoRA methods reach $99.67\%$ and $98.00\%$, respectively, with both methods achieving an AUC of $1.000$.
These results indicate that \Method is effective in backdoor detection on models trained with different PEFT methods.

\begin{table}[t]
\centering
\caption{Effectiveness of \Method on different transformer-based architecture models.}
\label{base_models_accuracy_auc}
\resizebox{0.48\textwidth}{!}{
\begin{tabular}{@{}ccc@{}}
\toprule
\textbf{Base Model} & \textbf{Detection Acc} & \textbf{Detection AUC} \\
\midrule
\centering
(Decoder-only) Llama-2-13B       & $99.67\%\pm 0.47\%$ & $1.000 \pm 0.000$ \\
(Decoder-only) Llama-3-8B        & $100.00\%\pm 0.00\%$ & $1.000 \pm 0.000$ \\
(Decoder-only) Llama-2-7B        & $100.00\%\pm 0.00\%$ & $1.000 \pm 0.000$ \\
(Decoder-only) Qwen1.5-7B-Chat   & $100.00\%\pm 0.00\%$ & $1.000 \pm 0.000$ \\
(Prefix Decoder-only) Chatglm-6B-v2  & $99.33\%\pm 0.47\%$ & $1.000 \pm 0.000$ \\
(Encoder-Decoder) Flan-t5-xl & $100.00\%\pm 0.00\%$ & $1.000 \pm 0.000$ \\
(Encoder-only) Roberta-base  & $98.33\%\pm 0.58\%$ & $1.000 \pm 0.000$ \\
\bottomrule
\end{tabular}
}
\end{table}

\mypara{Different Base Models}
We then investigate if \Method is effective for different base models. 
The corresponding results are summarized in \Cref{base_models_accuracy_auc}, which are evaluated on the datasets trained on \textit{toxic-backdoors-hard} using LoRA.
We observe that, for all base models, the backdoor detection AUC is $1.000$. 
Specifically, for Llama-3-8B, Llama-2-7B, Qwen1.5-7B-Chat, and Flan-t5-xl adapters, the detection accuracy reaches $100\%$.
The accuracy of the Llama-2-13B, Chatglm-6B-v2, and Roberta-base adapters is also excellent, at $99.67\%$, $99.33\%$, and $98.33\%$, respectively.
In summary, we have the following insights.

\begin{enumerate}
    
\item Based on the performances across the Llama series (with model sizes of 7B, 8B, and 13B), we demonstrate the effectiveness of \Method on different parameter scales.
   
\item The models in \Cref{base_models_accuracy_auc} can be categorized into pre-trained models (e.g., Llama-2-13B) and fine-tuned chat models (e.g., Qwen1.5-7B-Chat). The result shows that \Method is effective for different kinds of LLMs.
    
\item As mentioned in \Cref{subsec:target_llm}, there are four transformer architectures of LLMs, including Encoder-only, Decoder-only, Prefix Decoder-only, and Encoder-Decoder. We also consider the three most common transformer attention mechanisms: Multi-Head Attention, Grouped Query Attention, and Multi-Query Attention. 
\Method demonstrates general performances across these model architectures and attention mechanisms, indicating its effectiveness for backdoor detection in transformer-based LLMs.
\end{enumerate}

\begin{table}[t]
\caption{Effectiveness of \Method on different target projection matrices.}
\label{projection}
\centering
\resizebox{0.48\textwidth}{!}{
\begin{tabular}{@{}cccc@{}}
\toprule
\textbf{Projection Matrix} & \textbf{Rank} & \textbf{Detection Acc} & \textbf{Detection AUC} \\
\midrule
$[\Delta_q]$ & $512$   &  $100.00\%\pm 0.00\%$ & $1.000 \pm 0.000$                  \\
$[\Delta_k]$   & $512$           &  $100.00\%\pm 0.00\%$ & $1.000 \pm 0.000$                  \\
$[\Delta_v]$                         & $512$           &  $100.00\%\pm 0.00\%$ & $1.000 \pm 0.000$                 \\
$[\Delta_q , \Delta_k]$                     & $256$           &  $100.00\%\pm 0.00\%$ & $1.000 \pm 0.000$                    \\
$[\Delta_q$, $\Delta_v]$                     & $256$           &  $100.00\%\pm 0.00\%$ & $1.000 \pm 0.000$                  \\
$[\Delta_q$, $\Delta_k$, $\Delta_v$, $\Delta_o]$             & $128$           & $100.00\%\pm 0.00\%$ & $1.000 \pm 0.000$                    \\
\bottomrule
\end{tabular}
}
\end{table}

\mypara{Different Target Projection Matrices}
As discussed in~\Cref{Transformation}, we fine-tune adapters primarily by focusing on $\Delta_q$ and $\Delta_v$. 
However, it is equally important to explore whether \Method can be applied to other weight matrices within the self-attention module, as these matrices have also proven to be effective in tuning strategies~\cite{LoRA}.
For instance, in adapter training on $\Delta_q$, $\Delta_k$, $\Delta_v$, and $\Delta_o$, the concatenation of matrices in the $l$-th layer can be represented as $\Delta^{(l)}_{\rm concat} = [\Delta^{(l)}_q, \Delta^{(l)}_k, \Delta^{(l)}_v, \Delta^{(l)}_o]$.

Based on the \textit{toxic-backdoors-alpaca} dataset, we prepare adapters that are trained via LoRA on the Llama-2-7B model.
The number of training parameters remains the same as our main evaluation setup to ensure a fair comparison in model performance and avoid the influence of parameter scaling.
As shown in \Cref{projection}, our \Method demonstrates robust performance across various target projection matrices of adapters, all achieving $100\%$ detection accuracy.

\begin{figure}[t]
\centering \includegraphics[width=0.4\textwidth]{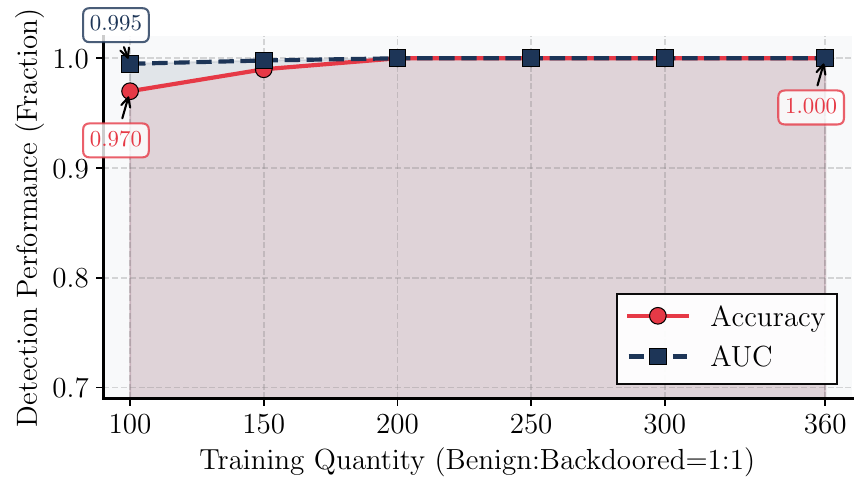}
\caption{Detection Performance across different training quantities.}
\label{fig:size}
\end{figure}

\mypara{Impact of Training Quantities}
\label{sec:size}
As mentioned in \Cref{section:Backdoor-detection}, our experiments are based on the most common $8:2$ ratio for training and testing set splits in deep learning, and $10\%$ of the training set is selected as the validation set. 
Here we also investigate the performance of \Method under different ratios of training datasets. 
We select the adapters trained via LoRA on the Llama-2-7B model within the \textit{toxic-backdoors-hard} datasets, keeping the testing data unchanged. 
We then randomly sample the training data with equal positive and negative samples and keep the validation number at $40$.
As shown in \Cref{fig:size}, \Method performs well across different data quantities ranging from $100$ to $360$.
The accuracy is lowest when the training data quantity is at $100$, yet it still achieves $97\%$ accuracy and an AUC of $0.995$.
When the training data quantity increases to $200$, \Method consistently reaches $100\%$ accuracy.

\begin{table}[t]
\centering
\caption{Performance of \Method with different architectures. (AdaptiveAvgPool refers to replacing the Convolutional Layer in \Cref{fig:net1} with an Adaptive Average Pooling Layer, while MaxPool indicates replacement with a Max Pooling Layer.)}
\label{tab:Archperformance}
\begin{tabular}{@{}ccc@{}}
\toprule
\textbf{Architecture}    & \textbf{Detection Accuracy}             & \textbf{Detection AUC}           \\ \midrule
\textbf{Original} & \textbf{$98.33\% \pm 0.47\%$}  & \textbf{$0.996\pm0.004$} \\  \midrule
AdaptiveAvgPool          & 50.00\% $\pm$ 0.00\%                   & 0.742 $\pm$ 0.035                \\
MaxPool                  & 89.33\% $\pm$ 1.25\%                   & 0.903 $\pm$ 0.035                \\
A1                       & 96.33\% $\pm$ 0.47\%                   & 0.973 $\pm$ 0.011                \\
A2                       & 98.00\% $\pm$ 0.816\%                   & 0.986 $\pm$ 0.013                \\ \bottomrule
\end{tabular}
\end{table}

\mypara{Explorations of \Method Architecture}
Referring to \Cref{classifier}, for $F_{input}$ with dimension $[64, 4096, 4096]$, removing the convolutional layer and directly using an MLP would lead to an extremely large first layer.
This occurs because the high-dimensional input lacks prior reduction, vastly increasing parameters and making the classifier hard to train and implement.
Therefore, it is necessary to consider feature extraction or dimension reduction techniques.
Here we first discuss the ablation study of dimensionality reduction strategy, where we replace the convolutional layer with a pooling layer.
Then we explore other efficient model architectures, demonstrating that the \Method framework is broadly adaptable and not limited to the default structure shown in \Cref{fig:net1}.

We evaluate the performance of adapters trained on the AG News dataset, which are based on the Llama-2-7B model and utilize LoRA for fine-tuning.
After replacing the convolutional layer with adaptive pooling layers, the detection performance is weaker, achieving only a $50\%$ accuracy.
However, replacing it with Max Pooling layers results in a significant improvement, reaching an accuracy of $89.33\%$.

To explore alternative structures for \Method classifier, we first increase the number of convolutional layers, incorporate adaptive pooling layers, and reduce the number of fully connected layers, resulting in architecture A1. 
Building on A1, we then replace the intermediate convolutional layers with residual layers, forming architecture A2 (refer to~\Cref{fig:subfigures}).
The comparable performance indicates that the original architecture is a suitable choice.

\begin{table}[t]
\centering
\caption{Effectiveness of \Method on various modality models.}
\label{Modalities}
\resizebox{0.48\textwidth}{!}{
\begin{tabular}{@{}cccc@{}}
\toprule
\textbf{Model}       & \textbf{Dataset}  & \textbf{Detection Acc} & \textbf{AUC} \\ \midrule
ViT-base              & CIFAR-10          &      $99.67\%\pm0.58\%$          &   $1.000\pm0.000$          \\
Qwen2-vl-2B          & VQAv2             &      $99.33\%\pm0.94\%$          &    $1.000\pm0.000$         \\ \bottomrule
\end{tabular}
}
\end{table}

\mypara{Extension Study Across Different Modalities}
Based on current experimental results, \Method performs well in detecting backdoor injections into LLMs using PEFT technology. 
Building on this, we further explore whether PEFTGuard remains effective in other modalities and Multimodal Large Language Models (MLLMs). 
Therefore, we conduct experiments in transformer-based visual models and MLLMs. 
Specifically, for the Visual Model (VM), we select Vit-base~\cite{dosovitskiy2020image} as the base model and, following the experimental setup of BadNets~\cite{BadNets}, use a white $5 \times 5$ pixel square in the bottom right corner of images from the CIFAR-10~\cite{krizhevsky2009learning} dataset as the trigger pattern. 
For the MLLM, we choose Qwen2-vl-2B~\cite{wang2024qwen2} as the base model, inserting the trigger phrase ``no cross, no crown'' in the textual modality of the VQAv2~\cite{goyal2017making} dataset, with the target output set to ``BOMB''.
As shown in \Cref{Modalities}, \Method maintains high detection performance in VM as well, achieving a detection accuracy of $99.67\%$, with $99.33\%$ in MLLM. 
This demonstrates that \Method also works for other modalities or MLLMs.

\subsection{Zero-Shot Transferability of \Method}
\label{sec:transferability}

We further investigate the zero-shot transferability of \Method across multiple aspects, including PEFT methods, LoRA ranks, and attacks.

\begin{table}[t]
    \caption{Transferability of \Method trained on the LoRA dataset. \colorbox{myblue!15}{\textbf{Blue}} indicates the training dataset.} 
    \label{diff_peft}
    \centering
    \begin{tabular}{@{}cccc@{}}
        \toprule
        \textbf{Method} & \textbf{Detection Acc} & \textbf{Detection AUC} \\ 
        \midrule
        \cellcolor{myblue!15}\textbf{LoRA}  & \cellcolor{myblue!15}$\mathbf{100.00\% \pm 0.00\%}$ & \cellcolor{myblue!15}$\mathbf{1.000\pm 0.000}$   \\ 
        QLoRA  & $100.00\% \pm 0.00\%$ & $1.000\pm 0.000$   \\ 
        LoRA+  & $99.00\% \pm 0.00\%$ & $1.000\pm 0.000$   \\ 
        DoRA  &  $99.33\% \pm 0.47\%$ & $1.000\pm 0.000$  \\
        AdaLoRA  & $50.00\% \pm 0.00\%$ & $0.000\pm 0.000$   \\
        \bottomrule
    \end{tabular}
\end{table}

\mypara{Transferability on Different PEFT Methods}
We aim to explore the zero-shot transferability of \Method on different PEFT methods. 
As indicated in \Cref{diff_peft}, the classifier trained on the LoRA is successfully transferred to QLoRA, LoRA+, and DoRA adapters, with detection accuracy exceeding $99\%$. 
While \Method demonstrates transferability across various PEFT methods, its transfer performance on AdaLoRA is weak. 
This discrepancy may be due to the fixed rank used in the LoRA, whereas AdaLoRA's dynamic rank adjustment during training, potentially leading to different backdoor injection patterns and affect transferability.

\begin{table}[t]
\caption{Transferability of \Method trained on LoRA rank $256$. \colorbox{myblue!15}{\textbf{Blue}} indicates the training dataset.}
\label{tab:lora_accuracy}
\centering
\begin{tabular}{cccc}
\toprule
\textbf{LoRA Rank} & \textbf{Detection Accuracy} & \textbf{Detection AUC} \\
\midrule
\cellcolor{myblue!15}$\mathbf{256}$ & \cellcolor{myblue!15}$\mathbf{100.00\% \pm 0.00\%}$ & \cellcolor{myblue!15}$\mathbf{1.000\pm 0.000}$ \\
$8$ & $98.67\% \pm 0.94\%$ & $1.000\pm 0.000$ \\
$16$ & $100.00\%\pm0.00\%$ & $1.000\pm 0.000$ \\
$32$ & $99.33\%\pm0.94\%$ & $1.000\pm0.000$ \\
$64$ & $100.00\%\pm0.00\%$ & $1.000\pm0.000$ \\
$128$ & $98.67\%\pm0.47\%$ & $0.998\pm0.003$ \\
$512$ & $98.99\%\pm0.82\%$ & $0.999\pm0.001$ \\
$1024$ & $98.67\%\pm0.47\%$ & $0.995\pm0.007$ \\
$2048$ & $96.00\%\pm1.63\%$ & $0.995\pm0.005$ \\
\bottomrule
\end{tabular}
\end{table}

\mypara{Transferability on Different Ranks of Adapters}
We also investigate the zero-shot transferability of \Method across different ranks of adapters, meaning that using the classifier trained on LoRA adapters with rank $256$ is transferred to other ranks under the same conditions.
\Cref{tab:lora_accuracy} shows that \Method exhibits excellent performance when transferred to LoRA ranks of less than $256$, achieving an average detection accuracy of more than $98.67\%$ and an average AUC of more than $0.998$. 
However, as the rank exceeds $256$, the detection performance declines, with the lowest average accuracy recorded at $96.00\%$.
This highlights the impact of LoRA rank on transferability.

\begin{table}[!t]
\centering
\renewcommand{\arraystretch}{1.2}
\caption{Zero-Shot transferability across different attacks. 
\colorbox{myblue!15}{\textbf{Blue}} indicates the training dataset.}
\label{tab:single_tranfer}
\resizebox{0.46\textwidth}{!}{
\begin{tabular}{llcccc}
\toprule
\multicolumn{2}{c}{\multirow{3}{*}{\makecell{\textbf{Detection}\\ \textbf{ACC / AUC}}}} 
 & \multicolumn{4}{c}{\textbf{Training Dataset}} \\
\cmidrule(lr){3-6}
\multicolumn{2}{c}{} 
 & \textbf{InsertSent} & \textbf{RIPPLES} & \textbf{Syntactic} & \textbf{StyleBkd} \\
\midrule
\multirow{7}{*}{\rotatebox[origin=c]{90}{\textbf{Transfer Dataset}}}
 & \textbf{InsertSent} 
   & \cellcolor{myblue!15}\makecell{$\mathbf{98.33\%}$\\/$\mathbf{0.996}$}
   & \makecell{$50.00\%$\\/$0.670$}
   & \makecell{$61.33\%$\\/$0.845$}
   & \makecell{$50.00\%$\\/$0.718$} \\
\cmidrule(lr){2-6}
 & \textbf{RIPPLES}
   & \makecell{$54.33\%$\\/$0.851$}
   & \cellcolor{myblue!15}\makecell{$\mathbf{100.00\%}$\\/$\mathbf{1.000}$}
   & \makecell{$56.67\%$\\/$0.828$}
   & \makecell{$50.00\%$\\/$0.795$} \\
\cmidrule(lr){2-6}
 & \textbf{Syntactic}
   & \makecell{$54.33\%$\\/$0.892$}
   & \makecell{$50.00\%$\\/$0.722$}
   & \cellcolor{myblue!15}\makecell{$\mathbf{99.33\%}$\\/$\mathbf{0.998}$}
   & \makecell{$50.00\%$\\/$0.849$} \\
\cmidrule(lr){2-6}
 & \textbf{StyleBkd}
   & \makecell{$64.33\%$\\/$0.965$}
   & \makecell{$50.00\%$\\/$0.662$}
   & \makecell{\textbf{$95.33\%$}\\ \textbf{/$0.986$}}
   & \cellcolor{myblue!15}\makecell{$\mathbf{100.00\%}$\\/$\mathbf{1.000}$} \\
\bottomrule
\end{tabular}}
\end{table}

\begin{table}[h!]
\centering
\caption{Zero-Shot transferability of \Method via contrastive learning and model fusion. (Notation: Sent = InsertSent, Word = RIPPLES, Syn = Syntactic, Sty = StyleBkd)}
\label{tab:model_fusion}
\resizebox{0.48\textwidth}{!}{
\begin{tabular}{@{}ccc@{}}
\toprule
\textbf{Model Fusion ($n=3$)}            & \makecell{\textbf{Known Attack} \\ \textbf{(Acc/AUC)}} & \makecell{\textbf{Unknown Attack} \\ \textbf{(Zero-Shot, Acc/AUC)}} \\ 
\midrule
Sent + Word + Syn       & $95.00\%$/$0.990$                   & (Sty) $95.00\%$/$0.989$                     \\
Sent + Word + Sty        & $100.00\%$/$1.000$                   & (Syn) $91.00\%$/$0.993$                    \\
Sent + Syn + Sty      & $94.00\%$/$0.999$                   & (Word) $90.00\%$/$0.946$                      \\
Word + Syn + Sty         & $93.00\%$/$0.966$                           & (Sent) $93.00\%$/$0.922$                                        \\
\bottomrule
\end{tabular}
}
\end{table}

\begin{table*}[h!]
\centering
\caption{Performance metrics under various adaptive attacks methods. For the C\&W attack, the detailed parameter settings can be found in Table \ref{c&w_parameters}.}
\label{adaptive_attacks}
\begin{tabular}{>{\centering\arraybackslash}p{4.5cm} >{\centering\arraybackslash}p{3.5cm} >{\centering\arraybackslash}p{2.5cm} >{\centering\arraybackslash}p{2.5cm} >{\centering\arraybackslash}p{2cm}}
\toprule
\multirow{2}{=}{\centering\arraybackslash \textbf{Attack Method}} & \multirow{2}{=}{\centering\arraybackslash \textbf{Parameters}} & \multicolumn{2}{c}{\textbf{Performance of Backdoored Model}} & \multirow{2}{=}{\centering\arraybackslash \textbf{ASR on PEFTGuard}} \\
 & & \textbf{CA under Attack} & \textbf{ASR under Attack} &  \\
\midrule
\multirow{1}{4.5cm}{\centering Initial Model \\} & - & $0.971 \pm 0.006$ & $0.999 \pm 0.007$ & $0\%\pm0\%$ \\
\midrule
\multirow{3}{4.5cm}{\centering Gaussian Noise \\(Scaled by Standard Deviation)} & scale=1 & $0.971 \pm 0.006$ & $0.999 \pm 0.007$ & $0\%\pm0\%$ \\
 & scale=3 & $0.947 \pm 0.079$ & $0.988 \pm 0.053$ & $0\%\pm0\%$ \\
 & scale=6 & $0.646 \pm 0.338$ & $0.793 \pm 0.379$ & $0\%\pm0\%$ \\
\multirow{3}{4.5cm}{\centering Gaussian Noise \\(Proportional to Parameter Size)} & parameter\_ratio=0.2 & $0.961 \pm 0.066$ & $0.978 \pm 0.140$ & $0\%\pm0\%$ \\
 & parameter\_ratio=0.4 & $0.920 \pm 0.163$ & $0.952 \pm 0.176$ & $0\%\pm0\%$ \\
 & parameter\_ratio=0.6 & $0.791 \pm 0.244$ & $0.962 \pm 0.152$ & $0\%\pm0\%$ \\
\midrule
\multirow{1}{4.5cm}{\centering FGSM} 
& $\epsilon = 1\times10^{-4}$ & $0.908 \pm 0.156$ & $0.843 \pm 0.305$ & $14\%\pm0\%$ \\
& $\epsilon = 1\times10^{-3}$ & $0.449 \pm 0.190$ & $0.139 \pm 0.099$ & $92\%\pm0\%$ \\
\midrule
\multirow{3}{4.5cm}{\centering I-FGSM} & $\epsilon = 1\times10^{-4}, \alpha = 1\times10^{-5}$ & $0.971\pm0.006$ & $0.983\pm0.071$ & $16\%\pm10\%$ \\
& $\epsilon = 1\times10^{-3}, \alpha = 1\times10^{-4}$ & $0.971\pm0.006$ & $0.791\pm0.354$ & $65\%\pm13\%$ \\
 & $\epsilon = 5\times10^{-3}, \alpha = 5\times10^{-4}$ & $0.969\pm0.008$ & $0.455\pm0.481$ & $100\%\pm0\%$ \\
\midrule
\multirow{3}{4.5cm}{\centering PGD} &  $\epsilon = 1\times10^{-4}, \alpha = 1\times10^{-5}$ & $0.971\pm0.006$ & $0.975\pm0.089$& $21\%\pm9\%$ \\
 & $\epsilon = 1\times10^{-3}, \alpha = 1\times10^{-4}$ & $0.971\pm0.006$ & $0.786\pm0.350$ & $65\%\pm13\%$ \\
 & $\epsilon = 5\times10^{-3}, \alpha = 5\times10^{-4}$ & $0.834\pm0.199$ & $0.408\pm0.331$ & $100\%\pm0\%$ \\
 \midrule
 \multirow{5}{4.5cm}{\centering C\&W} & $P_1$ & $0.971\pm0.006$ &$0.940\pm0.160$ & $34\%\pm8.5\%$ \\
  & $P_2$ & $0.734\pm0.173$ & $0.082\pm0.168$ & $83.5\%\pm2.2\%$ \\
  & $P_3$ & $0.971\pm0.008$ & $0.611\pm0.372$ & $67.0\%\pm3.0\%$\\
  & $P_4$ & $0.000\pm0.000$ & $0.000\pm0.000$ &$70.8\%\pm2.0\%$\\
  & $P_5$ & $0.970\pm0.008$ & $0.570\pm0.357$ & $86.8\%\pm1.2\%$\\
\bottomrule
\end{tabular}
\end{table*}

\mypara{Transfer Across Different Attacks}
We aim to explore whether \Method exhibits Zero-Shot Transferability from known attacks (those the model has been trained on) to the detection of unknown attacks (those the model has not encountered), using adapters trained on four different attack methods on AG News as shown in Table \ref{table:Dataset}.
Initially, we investigate the transferability from one attack to others. 
From \Cref{tab:single_tranfer}, we find that only the classifier trained on Syntactic attacks could transfer well to StyleBkd, achieving a detection accuracy of $95.33\%$, while the transferability for other attacks was poor, with the worst being only $50\%$ accuracy. 
However, we also discover that despite the poor accuracy, the AUC indicated that \Method possesses some potential for zero-shot transfer.

To enhance the zero-shot transferability across different attacks, we employ supervised contrastive loss for contrastive learning~\cite{khosla2020supervised}, training on three different attack datasets. 
Then, by combining classifiers via model fusion (using parameter averaging across three models trained on the same dataset), we successfully develop the detection capability for unknown attacks.
As shown in \Cref{tab:model_fusion}, by leveraging contrastive learning and model fusion, \Method, after being trained on any three attack datasets, demonstrates better zero-shot detection capability on unknown attacks.
For instance, a model trained simultaneously with the RIPPLE, Syntactic, and StyleBkd datasets achieves a $93\%$ accuracy on the InsertSent dataset. 
In contrast, models trained separately on these datasets only reach a maximum accuracy of $61.33\%$ when transferred to InsertSent.

More details of the ablation study are discussed in \Cref{sec:ablation_trans} in the Appendix.

\subsection{Adaptive Attacks}

\mypara{Motivation}
We now consider a real-world scenario where the adversary aims to bypass the potential detector using adaptive attacks.
We consider the adversary's goal to be disrupting \Method model through the introduction of noise perturbations. 
Attackers can directly add Gaussian noise or utilize our publicly available \Dataset to train their own classifier and evade detection based on classifiers by adjusting the weights of the adapters.

\mypara{Adaptive Attacking Scenarios}
Therefore, we discuss the adversary's use of different perturbations to attack \Method classifier, including the addition of Gaussian noise, FGSM, I-FGSM, PGD, and C\&W methods, with the adapters trained on IMDB based on the Llama-2-7B model with LoRA.
As shown in \Cref{adaptive_attacks}, we consider two methods of adding Gaussian noise.
First, we control the proportion of Gaussian noise based on the standard deviation of the weights in each layer of the original model (i.e., scaled by standard deviation).
Secondly, we set the standard deviation of the noise equal to the standard deviation of the model's original weights (scale = $5$), and adjust the proportion of Gaussian noise added within the total parameters of the model (i.e., proportional to parameter size). 
For optimization-based adversarial attacks, we assume that the adversary can optimize the adversarial examples based on their trained model and transfer them to our \Method model.
Note that here we consider a strong adversary that can train another detection model using the same training dataset and hyperparameters as \Method.

\mypara{Results Analysis}
As shown in \Cref{adaptive_attacks}, as the proportion of noise increases, both methods affect the performance of the backdoored model, reducing both CA and ASR, but the ASR on \Method classifier remains at $0\%$, indicating that \Method is robust against Gaussian noise.
For FGSM, when $\epsilon = 1\times 10^{-3}$, although the ASR on \Method can reach $92\%$, the CA of the backdoored model drops from $97.1\%$ to $44.9\%$, and the ASR also decreases from 0.999 to 0.139. 
The same conclusion can be seen in I-FGSM, PGD, and C\&W.
We also observe that when the intensity of the attack increases, the ASR on \Method can rise to $100\%$ in I-FGSM and PGD and $86.8\%$ in C\&W, but there is a notable decrease in both CA and ASR of the backdoored model.
This suggests that even if the adversary can adjust the adapter's weights to evade detection by \Method by training their own classifier, such adjustments significantly degrade the overall performance of the model, representing a considerable cost to the adversary.

\begin{table}[t]
\centering
\caption{Performance of backdoor mitigation methods.}
\label{defense}
\begin{tabular}{ccc}
\toprule
\textbf{Method}         & \textbf{ASR~$(\downarrow)$}    & \textbf{Accuracy~($\uparrow$)} \\ \midrule
Original Model & $100.00\%$  & $96.88\%$  \\
SFT  & $9.80\%$ & $93.92\%$  \\
DPO  & $0.00\%$    & $64.56\%$       \\
Fine-mixing & $7.20\%$ & $96.12\%$  \\ \bottomrule
\end{tabular}
\end{table}

\subsection{Backdoor Mitigation}

To mitigate the backdoor, we consider three methods: SFT, DPO, and Fine-mixing. 
Our target defense adapter is trained on the IMDB dataset.
As shown in~\Cref{defense}, the original backdoored model has an ASR of $100.00\%$ and a CA of $96.88\%$. 
The worst-performing method is DPO, which eliminates the backdoor to $0.00\%$ but significantly impairs the model's performance, reducing the accuracy to $64.56\%$.
Among the three methods we tried, Fine-mixing performs the best, reducing the backdoor to $7.20\%$ while only decreasing the accuracy on the clean dataset to $93.92\%$, compared with the $4.60\%$ of no attack ASR.
This suggests that Fine-mixing serves as a good defense to remove backdoors. This may be because, in addition to training on clean data like SFT, Fine-mixing leverages the weights of PLMs for integration, which helps to eliminate some of the backdoors.
Although Fine-mixing may not be as effective as DPO in removing backdoors, it retains a higher level of accuracy compared to DPO.

\section{Limitations}
\label{sec:limitations}

\mypara{Transferability}
Since \Method relies on training a meta-classifier to detect backdoored adapters, any variation in input dimensions requires training a new classifier. This constraint limits the zero-shot transferability of \Method across different LLMs.
Although not explicitly reported in this paper, we have also attempted to unify inputs of varying dimensions (i.e., different LLMs) by downsampling.
Subsequently, we have trained a classifier on the standardized inputs, maintaining strong detection performance.
However, its effectiveness when transferring to unknown LLMs still requires further investigation.
Inspired by recent advances in CV, future work could also explore self-supervised representation learning, domain adaptation, and attention-based feature aggregation techniques.

\mypara{Practical Deployment}
Although \Method is effective, when applied to a completely new detection scenario, it requires substantial time for training and deployment.
However, as shown in \Cref{fig:size}, we can achieve strong detection performance with only $100$ training samples, which can reduce deployment time. 
In addition, to further reduce training overhead and improve transferability, future work may explore leveraging \Dataset samples for domain adaptation, model distillation, and related techniques.

\mypara{Backdoor Pattern Explanation} 
In our work, we take a step toward explaining backdoor patterns in the model parameters as shown in \Cref{fig:feature}, which shows a clear distinction between backdoored and benign adapters. Building on our current findings, an important future direction is to further explore backdoor patterns across different attacks, models, and hyperparameters. In particular, analyzing these patterns in the task vector space could be promising, and prior works in the CV domain~\cite{mazeika2023trojan,langosco2023detecting} may provide helpful insights.

\section{Related Work}

\mypara{LLM} 
Recently, LLMs have achieved great success in  NLP domain~\cite{zhao2023survey, minaee2024large}.
Trained on large amounts of text, LLMs have developed strong language modeling capabilities and assisted humans in solving complex tasks, such as OpenAI's ChatGPT~\cite{openai2023gpt4} and GPT-4~\cite{chatgpt} and Microsoft's Copilot systems~\cite{microsoft_copilot}. There are many open-source LLMs (such as Llama-3~\cite{metallama3}, Mixtral~\cite{jiang2023mistral} and Qwen~\cite{bai2023qwen}), and it's necessary to fine-tune models for specific downstream tasks.

\mypara{PEFT}
PEFT is an excellent fine-tuning method that reduces resource consumption while maintaining performance comparable to full-parameter fine-tuning.
LoRA~\cite{LoRA} is the most representative parameter-efficient fine-tuning mechanism widely adopted for LLMs. Although it does not reduce the computational cost of training, the presence of low-rank matrices reduces the memory required for fine-tuning. 
QLoRA~\cite{QLoRA} employs quantization techniques to optimize LoRA's storage and computational efficiency, further reducing memory usage and computational cost while maintaining model performance. 
AdaLoRA~\cite{AdaLoRA} considers the varying importance of pre-trained weights across different layers and automatically adjusts the rank of low-rank matrices based on the importance scores of weight matrices to further reduce training parameters. 
LoRA+~\cite{LoRA+} uses different learning rates for low-rank matrices A and B to improve performance and fine-tuning speed. 
DoRA~\cite{DoRA} decomposes the pre-trained weight into magnitude and direction components for fine-tuning, thus enhancing the fine-tuning performance.

\mypara{Backdoor Attacks}
In the field of NLP, researchers have studied various backdoor attacks.
Inserting triggers into the data is the most common and effective attack method~\cite{RIPPLES, InsertSent, yang-etal-2021-careful} leads models to malicious behaviors. 
More seriously, injecting backdoors into the training of LLMs can cause models to generate toxic responses when triggered, leading to severe consequences~\cite{wan2023poisoning, zhang2024instruction, yan2023backdooring}, and can also cause significant security issues in applications based on LLMs~\cite{yang2024watch}.
This raises concerns about open-source LLMs on the internet. In addition, due to the effectiveness of PEFT, many people share their PEFT adapter models online to accomplish various downstream tasks, which may result in the spread of backdoors~\cite{liu2024loraasanattack}.
Besides, Dong et al.~\cite{dong2023philosopher} also propose two novel backdoor attacks targeting adapters, POLISHED and FUSION, which successfully manipulate the LLMs to perform malicious actions.
Therefore, backdoor detection methods for PEFT are highly demanded.

\mypara{Backdoor Detection on LLMs}
Early methods for backdoor detection in NLP tasks include
Trojan-Miner~\cite{azizi2021t} targets DNN-based text classification tasks, using a seq-to-seq model to probe suspicious classifiers and generate sentences that may contain trojan triggers to detect the backdoors.
AttenTD~\cite{lyu-etal-2022-study} is an attention-based detector that provides a set of neutral trigger candidates and distinguishes backdoored models through attention anomalies.
PICCOLO~\cite{liu2022piccolo} determines the presence of backdoors by analyzing the model's sensitivity to trigger words. 
Note that one of the most similar methods as PEFTGuard is MNTD~\cite{xu2021detecting}, which relies on representation vectors of model outputs, whereas \Method directly leverages the model parameters.
Besides, Zeng et al.~\cite{DBLP:conf/ndss/ZengCP0DJ25} propose the first framework, CLIBE, for detecting dynamic backdoors in transformer-based NLP models by introducing few-shot perturbations into the suspect model's parameters.
However, with the rise of decoder-only architectures in LLMs~\cite{brown2020language,openai2023gpt4}, there is an increasing focus on tasks involving the generation of coherent content.
The main behavior of backdoor attacks on this task is typically to induce the LLMs to generate incorrect or toxic text outputs, which brings new challenges to traditional detection methods.
Specifically, trigger generation requires optimizing text triggers through changes in classification labels, and optimization-based trigger inversion methods struggle to generate precise triggers from changes in the discrete output domain in generation tasks. 
Moreover, attention analysis methods that rely on preset trigger words face challenges because adding a new word can shift the model's attention to fit the context, making detection much more challenging.

\section{Conclusion}

In this paper, we conduct the first in-depth and comprehensive analysis to reveal the security vulnerabilities of backdoored PEFT-based adapters.
To promote the development of backdoor detection against adapters, we construct the first dataset, \Dataset, which contains various backdoored or benign adapters. 
Meanwhile, we propose the first backdoor detection method, named \Method. 
It does not require any additional information during the detection, only access to the PEFT adapter parameters.
\Method achieves state-of-the-art performance in detecting various types of adapters, which are generated from different datasets, backdoor attacks, PEFT methods, and various base LLMs.
In addition, \Method demonstrates a zero-shot transferability across different PEFT methods, adapter ranks, and backdoor attacks. 
Furthermore, we show that \Method is robust against different adaptive attacks.
Overall, we hope that the proposed \Dataset and \Method will play a key role in advancing the security governance of adapters in the open-source platforms.

\section{Acknowledgments}

We sincerely thank the reviewers and our shepherd for their constructive feedback, which improved the quality of this work.
This work is supported by the National Natural Science Foundation of China (No.62425205, No.62376210, and No.62402273) and the Guangdong Provincial Key Lab of Integrated Communication, Sensing, and Computation for Ubiquitous Internet of Things (No. 2023B1212010007).

\clearpage
\bibliographystyle{plain}
\bibliography{example.bib}

\clearpage
\appendix

\section{Ablation Study about Zero-shot Transferability on Unknown Attacks}
\label{sec:ablation_trans}

\begin{table}[ht!]
\centering
\caption{Ablation study performance of Zero-Shot transferability analysis. (Notation: Sent = InsertSent, Word = RIPPLES, Syn = Syntactic, Sty = StyleBkd)}
\label{tab:transferability_ablation}
\resizebox{0.48\textwidth}{!}{
\begin{tabular}{@{}ccccr@{}}
\toprule
Training Attack & Fusion (models) & CL & Detection Acc (\%) & AUC \\ 
\midrule
Sent+Sty        & \checkmark, $3$   & \checkmark & $59.00$ & $0.973$ \\
Sent+Word       & \checkmark, $3$   & \checkmark & $50.00$ & $0.815$ \\
Word+Sty        & \checkmark, $3$   & \checkmark & $57.00$ & $0.971$ \\
Sent+Word+Sty   & -               & \checkmark & $68.00$ & $0.972$ \\
Sent+Word+Sty   & \checkmark, $3$   & -          & $74.00$ & $0.936$ \\
Sent+Word+Sty   & \checkmark, $2$   & \checkmark & $85.00$ & $0.988$ \\
Sent+Word+Sty   & \checkmark, $3$   & \checkmark & $91.00$ & $0.993$ \\
\bottomrule
\end{tabular}
}
\end{table}

In \Cref{sec:transferability}, we employ Contrastive Learning (CL), combined with model fusion via parameter averaging, enabling models trained on three different attack datasets to generalize to another unknown attack dataset.
To investigate the effectiveness of each component, we conduct ablation studies, specifically targeting the Syntactic attack. 
This is because the transfer effectiveness to Syntactic is consistently the worst among other attacks (Shown in~\Cref{tab:single_tranfer}).

From~\Cref{tab:transferability_ablation}, we can observe the impact of varying numbers of attacks, the number of fusions, and CL on the outcomes.
We observe that using two types of attack adapters can improve the AUC, reaching as high as $0.973$ but with a low DA $59\%$.
Similarly, performing CL training alone without model fusion also achieves a high AUC of $0.973$, but the DA only reaches $68\%$. 
Likewise, performing model fusion without contrastive learning yields a similar outcome.
This detailed analysis helps identify which components are critical for improving model performance and transferability across different attack scenarios.

\section{Related Work on Backdoor Mitigation Methods}
\label{sec:related_mitigation}

Backdoor mitigation methods aim to directly eliminate backdoors from models and can be combined with detection techniques to first identify and then remove them.
We focus on methods to eliminate backdoors through training techniques~\cite{yao2019latent, sha2022fine, DPO, fine-mixing}.
Experiments by Yao et al.~\cite{yao2019latent} and Sha et al.~\cite{sha2022fine} both indicate that fine-tuning backdoored models on a clean subset of training samples can mitigate the backdoors.
Rafailov et al.~\cite{DPO} propose Direct Preference Optimization (DPO), an optimization method specifically designed for LLMs. 
This method utilizes the mapping relationship between reward functions and optimal policies, demonstrating that this constrained reward maximization problem can be accurately optimized through single-stage policy training. 
By setting texts that contain backdoor triggers but have normal answers as preferred texts, the DPO method can effectively eliminate backdoor influences in the model.
Zhang et al.~\cite{fine-mixing} propose the Fine-mixing method, which considers the clean pre-trained model weights before fine-tuning on clean data and mixes the backdoored weights with clean pre-trained weights. In addition, they utilize Embedding Purification (E-PUR) to detect and eliminate potential backdoor techniques within embeddings.

\begin{figure}[h!]
    \centering
    \begin{subfigure}{\linewidth}
        \centering
        \includegraphics[width=0.45\linewidth]{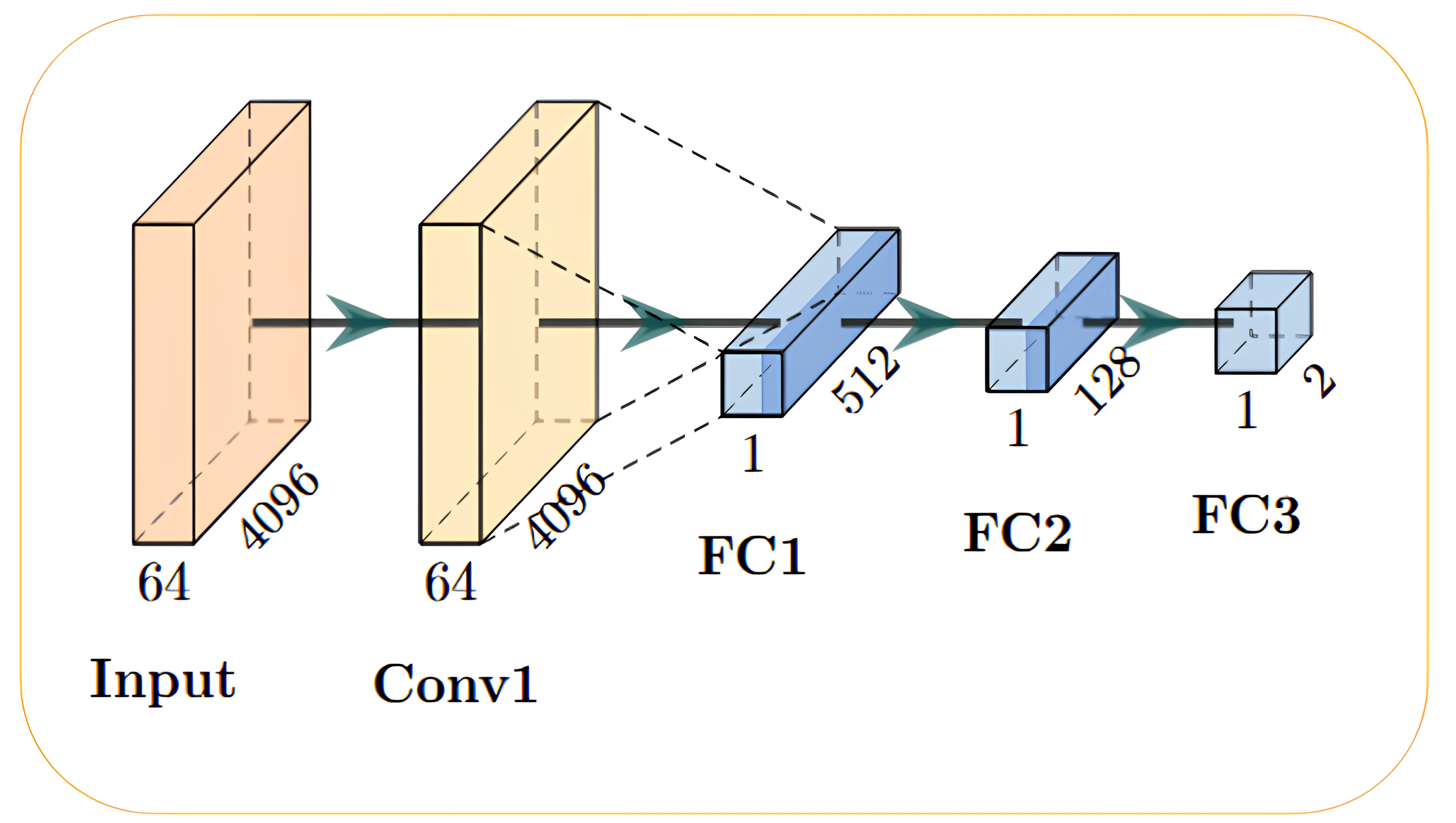}
        \caption{Architecture of $\mathsf{PEFTGuard}_{\rm cls}$.}
        \label{fig:net1}
    \end{subfigure}
    
    \vspace{1em}

    \begin{subfigure}{\linewidth}
        \centering
        \includegraphics[width=0.7\linewidth]{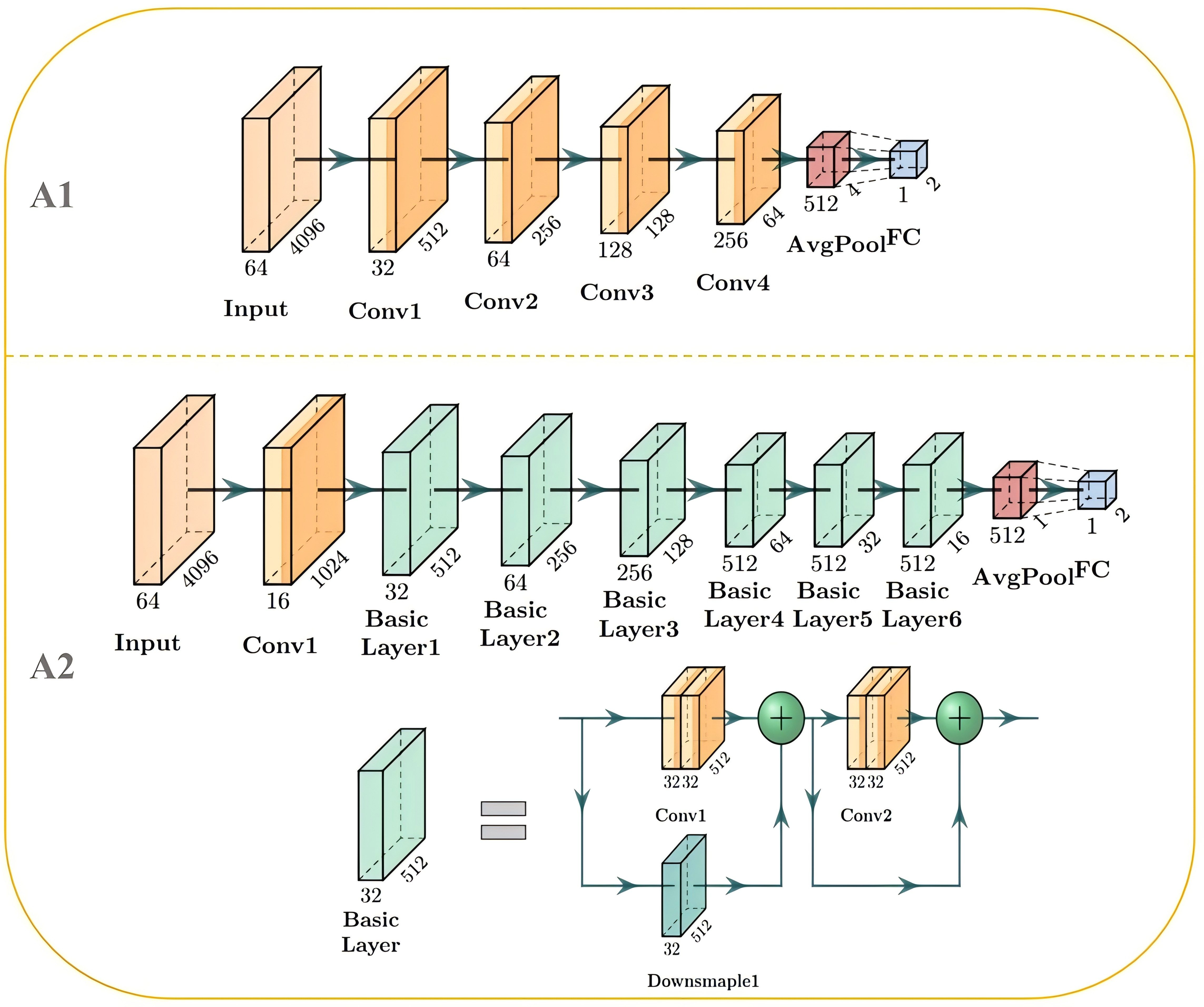}
        \caption{Alternative Architectures of \Method, representing two examples among many potential configurations.}
        \label{fig:subfigures}
    \end{subfigure}
    
    \caption{Illustration of \Method architectures.}
    \label{fig:combined}
\end{figure}

\begin{table*}[h!]
    \caption{Details of our \Dataset. ``-'' indicates empty because the instruction-following datasets can't be evaluated for clean accuracy. ``*'' indicates that in IMDB and AG News datasets, the benign adapters remain the same across different attacks in the same training dataset.(Unless specified otherwise, the target projection matrices in the PEFT method are, by default, applied to the query and value matrices.)}
\label{table:Dataset}
\centering
\resizebox{\textwidth}{!}{
\begin{tabular}{cccccccccc}
    \toprule
    \multirow{2}{*}{\textbf{Base Model}} & \multirow{2}{*}{\textbf{Dataset}} & \multirow{2}{*}{\textbf{Attack Method}} & \multirow{2}{*}{\textbf{PEFT Method}} & \multirow{2}{*}{\textbf{Rank}} & \multicolumn{1}{c}{\textbf{Number}} & \multicolumn{2}{c}{\textbf{Clean Accuracy}} & \multirow{2}{*}{\textbf{ASR}} \\
    \cmidrule(lr){6-6} \cmidrule(lr){7-8}
    & & & & & \textbf{Benign/Backdoored} & \textbf{Benign} & \textbf{Backdoored} & \\
    \midrule
    \multirow{1}{*}{Llama-2-7B} & SQuAD & InsertSent & LoRA & $256$ & $250/250$ & $0.647$ & $0.661$ & $0.997$ \\
    \midrule
    \multirow{1}{*}{Llama-2-7B} & toxic-backdoors-alpaca & Word & LoRA & $256$ & $250/250$ & - & - & $0.964$ \\
    \midrule
    \multirow{4}{*}{Llama-2-7B} & \multirow{4}{*}{IMDB} & RIPPLES & LoRA & $256$ & $250^*/250$ & $0.970$ & $0.973$ & $0.943$ \\
    & & InsertSent & LoRA & $256$ & $250^*/250$ & $0.970$ & $0.970$ & $0.998$ \\
    & & Syntactic & LoRA & $256$ & $250^*/250$ & $0.970$ & $0.969$ & $0.987$ \\
    & & StyleBkd & LoRA & $256$ & $250^*/250$ & $0.970$ & $0.960$ & $0.949$ \\
    \midrule
    \multirow{4}{*}{Llama-2-7B} & \multirow{4}{*}{AG News} & RIPPLES & LoRA & $256$ & $250^*/250$ & $0.940$ & $0.940$  & $0.948$ \\
    & & InsertSent & LoRA & $256$ & $250^*/250$ & $0.940$ & $0.938$ & $0.969$ \\
    & & Syntactic & LoRA & $256$ & $250^*/250$ & $0.940$ & $0.940$ & $0.986$ \\
    & & StyleBkd & LoRA & $256$ & $250^*/250$ & $0.940$ & $0.943$ & $0.927$ \\
    \midrule
    \multirow{12}{*}{Llama-2-7B} & \multirow{12}{*}{toxic-backdoors-hard} 
    & Sentence & LoRA & $256$ & $250/250$ & - & - & $0.926$ \\
    & &Sentence & LoRA (q) & $512$ & $250/250$ & - & - & $0.938$ \\
    & &Sentence & LoRA (k) & $512$ & $250/250$ & - & - & $0.936$ \\
    & &Sentence & LoRA (v) & $512$ & $250/250$ & - & - & $0.942$ \\
    & &Sentence & LoRA (q,k) & $256$ & $250/250$ & - & - & $0.934$ \\
    & &Sentence & LoRA (q,k,v,o) & 128 & $250/250$ & - & - & $0.979$ \\
    & & Sentence & QLoRA & $256$ & $250/250$ & - & - & $0.796$ \\
    & & Sentence & DoRA & $256$ & $250/250$ & - & - & $0.787$ \\
    & & Sentence & LoRA+ & $8$ & $250/250$ & - & - & $0.644$ \\
    & & Sentence & AdaLoRA & $8$ & $250/250$ & - & - & $0.112$ \\
    & & Sentence & LoRA & $8$ & $50/50$ & - & - & $0.585$ \\
    & & Sentence & LoRA & $16$ & $50/50$ & - & - & $0.707$ \\
    & & Sentence & LoRA & $32$ & $50/50$ & - & - & $0.739$ \\
    & & Sentence & LoRA & $64$ & $50/50$ & - & - & $0.816$ \\
    & & Sentence & LoRA & $128$ & $50/50$ & - & - & $0.734$ \\
    & & Sentence & LoRA & $512$ & $50/50$ & - & - & $0.846$ \\
    & & Sentence & LoRA & $1024$ & $50/50$ & - & - & $0.833$ \\
    & & Sentence & LoRA & $2048$ & $50/50$ & - & - & $0.814$ \\
    \midrule
    \multirow{1}{*}{Llama-2-13B} & toxic-backdoors-hard & Sentence & LoRA & $256$ & $250/250$ & - & - & $0.835$ \\
    \midrule
    \multirow{1}{*}{Llama-3-8B} & toxic-backdoors-hard & Sentence & LoRA & $256$ & $250/250$ & - & - & $0.843$ \\
    \midrule
    \multirow{1}{*}{Qwen1.5-7B-Chat} & toxic-backdoors-hard & Sentence & LoRA & $256$ & $250/250$ & - & - & $0.677$ \\
    \midrule
    \multirow{1}{*}{ChatGLM-6B-v2} & toxic-backdoors-hard & Sentence & LoRA & $256$ & $250/250$ & - & - & $0.641$ \\
    \midrule
    \multirow{1}{*}{flan-t5-xl} & toxic-backdoors-hard & Sentence & LoRA & $256$ & $250/250$ & - & - & $0.479$ \\
    \midrule        
    \multirow{1}{*}{Roberta-base} & IMDB & InsertSent & LoRA & $256$ & $250/250$ & $0.955$ & $0.950$ & $1.000$  \\
    \midrule
    \multirow{1}{*}{Qwen2-vl-2B} & VQAv2 & InsertSent & LoRA & $16$ & $250/250$ & $0.735$ & $0.738$ & $0.649$  \\
    \midrule
    \multirow{1}{*}{ViT-base} & CIFAR-10 & BadNets & LoRA & $16$ & $250/250$ & $0.985$ & $0.984$ & $0.921$ \\
    \bottomrule
\end{tabular}
}
\end{table*}

\begin{table}[h]
\centering
\caption{Summary of backdoor injection datasets and tasks.} 
\label{tab:datasets_tasks}
\resizebox{0.48\textwidth}{!}{
\begin{tabular}{@{}ccc@{}}
\toprule 
\textbf{Type} & \textbf{Dataset} & \textbf{Task} \\ 
\midrule
\multirow{5}{*}{Task-Specific} & SQuAD~\cite{rajpurkar2016squad} & Question Answering (QA) \\
                               & AG News~\cite{AG_news} & Topic Classification  \\
                               & IMDB Movie~\cite{IMDB} & Sentiment Classification (SC) \\
                               & CIFAR-10~\cite{krizhevsky2009learning} & Image Classification \\
                               & VQAv2~\cite{goyal2017making} & Visual Question Answering \\
\midrule
\multirow{2}{*}{Instruction-Following (IF)} & toxic-backdoors-hard~\cite{toxic_backdoors_hard} & Generation \\
                                       & toxic-backdoors-alpaca~\cite{toxic_backdoors_alpaca} & Generation \\
\bottomrule
\end{tabular}
}
\end{table}

\begin{table}[h!]
\centering
\caption{Parameter settings for the C\&W attack.}
\label{c&w_parameters}
\resizebox{0.48\textwidth}{!}{
\begin{tabular}{@{}cc@{}}
\toprule
\textbf{Parameter Set} & \textbf{Settings} \\
\midrule
$P_1$ & $c=1 \times 10^{-4}$, $\kappa=0$, $\text{iter}=20$, $\text{lr}=1 \times 10^{-5}$ \\
$P_2$ & $c=5 \times 10^{-3}$, $\kappa=0$, $\text{iter}=20$, $\text{lr}=5 \times 10^{-4}$ \\
$P_3$ & $c=0.1$, $\kappa=0$, $\text{iter}=30$, $\text{lr}=1 \times 10^{-4}$ \\
$P_4$ & $c=0.1$, $\kappa=5$, $\text{iter}=30$, $\text{lr}=1 \times 10^{-4}$ \\
$P_5$ & $c=0.5$, $\kappa=10$, $\text{iter}=30$, $\text{lr}=1 \times 10^{-4}$ \\
\bottomrule
\end{tabular}
}
\end{table}

\end{document}